\documentclass[twocolumn]{autart}    % Enable this line and disable the 
\usepackage{graphicx}          % Include this line if your 
\usepackage{enumitem} % To reference numbered items
\usepackage{bbm} % This is for \mathbb for lower case letters
\usepackage{color}   %For coloured text. For some reason, autart.cls does not have this predefined.
\usepackage{amsmath} % assumes amsmath package installed
\usepackage{amssymb}  % assumes amsmath package installed
\usepackage{amsfonts}
\usepackage{arydshln} %Dashed lines in matrices
\usepackage{balance} %balance last page

\usepackage{subcaption}

\usepackage{etoolbox} %Used to create border matrix code
\let\bbordermatrix\bordermatrix
\patchcmd{\bbordermatrix}{8.75}{4.75}{}{}
\patchcmd{\bbordermatrix}{\left(}{\left[}{}{}
\patchcmd{\bbordermatrix}{\right)}{\right]}{}{}

\newcommand{\vect}[1]{\boldsymbol{#1}}
\newcommand{\mat}[1]{\boldsymbol{#1}}
\newcommand{\wt}[1]{\widetilde{#1}}
\newcommand{\wh}[1]{\widehat{#1}}

\renewcommand{\eqref}[1]{Eq.~(\ref{#1})}  %Modified equation reference

\newtheorem{remark}{Remark}
\newtheorem{theorem}{Theorem}
\newtheorem{lemma}{Lemma}

\newtheorem{corollary}{Corollary}

\newtheorem{assumption}{Assumption}

\begin{document}

\begin{frontmatter}
%\runtitle{Insert a suggested running title}  % Running title for regular 
                                              % papers but only if the title  
                                              % is over 5 words. Running title 
                                              % is not shown in output.

\title{An Influence Network Model to Study Discrepancies in Expressed and Private Opinions\thanksref{footnoteinfo}} % Title, preferably not more 
                                                % than 10 words.

\thanks[footnoteinfo]{This paper was not presented at any IFAC 
meeting. Corresponding author: M. Ye.}

\author[anu]{Mengbin Ye}\ead{m.ye@rug.nl},    % Add the 
\author[rug]{Yuzhen Qin}\ead{y.z.qin@rug.nl},               % e-mail address 
\author[rug]{Alain Govaert}\ead{a.govaert@rug.nl},  % (ead) as shown
\author[anu,hdu,data61]{Brian D.O. Anderson}\ead{brian.anderson@anu.edu.au},
\author[rug]{Ming Cao}\ead{m.cao@rug.nl}

\address[anu]{Research School of Engineering, the Australian National University, Canberra, A.C.T. 2601, Australia}  % Please supply                                              
\address[rug]{Faculty of Mathematics and Natural Sciences, University of Groningen, The Netherlands}             % full addresses
\address[hdu]{School of Automation, Hangzhou Dianzi University, Hangzhou 310018, China} 
\address[data61]{Data61-CSIRO, Canberra, A.C.T. 2601, Australia} 
          
                             % keyword list or with the 
                                          % help of the Automatica 
                                          % keyword wizard

\begin{abstract}                          % Abstract of not more than 200 words.
In many social situations, a discrepancy arises between an individual's private and expressed opinions on a given topic. Motivated by Solomon Asch's seminal experiments on social conformity and other related socio-psychological works, we propose a novel opinion dynamics model to study how such a discrepancy can arise in general social networks of interpersonal influence. Each individual in the network has \emph{both a private and an expressed opinion}: an individual's private opinion evolves under social influence from the expressed opinions of the individual's neighbours, while the individual determines his or her expressed opinion under a pressure to conform to the average expressed opinion of his or her neighbours, termed the \emph{local public opinion}. General conditions on the network that guarantee exponentially fast convergence of the opinions to a limit are obtained. Further analysis of the limit yields several semi-quantitative conclusions, which have insightful social interpretations, including the establishing of conditions that ensure every individual in the network has such a discrepancy. Last, we show the generality and validity of the model by using it to explain and predict the results of Solomon Asch's seminal experiments.
\end{abstract}

\end{frontmatter}

\section{Introduction}
The study of dynamic models of opinion evolution on social networks has recently become of interest to the systems and control community. Most models are agent-based, in which the opinion(s) of each individual (agent) evolve via interaction and communication with neighbouring individuals.
%; such models share many similarities and parallels with works on coordination of multi-agent systems. 
%Models of opinion evolution have also greatly interested researchers from sociology, computer science, and physics, among others. 
This paper aims to develop a novel opinion dynamics model as a general theoretical framework to study how discrepancies arise in individuals' private and expressed opinions, and thus bridge the current gap between socio-psychological studies on conformity and dynamic models of interpersonal influence. Interested readers are referred to \cite{proskurnikov2017tutorial,proskurnikov2018tutorial_2,flache2017opdyn_survey} for surveys on the many works on opinion dynamics models.

Discrepancies in private and expressed opinions of individuals can arise in many situations, with a variety of consequential phenomena. Over one third of jurors in criminal trials would have privately voted against the final decision of their jury \cite{waters2009jury}. Large differences between a population's private and expressed opinions can create discontent and tension, a factor associated with the Arab Spring movement \cite{goodwin2011arabspring} and the fall of the Soviet Union \cite{kuran1989revolutions}. Access to the public action of individuals, without being able to observe their thoughts, can create informational cascades where all subsequent individuals select the wrong action \cite{bikhchandani1992informational_cascade}.  
%Interest in such discrepancies has been sustained over the years, in no small part due to the wide range of scenarios in which it can be encountered (the above gives but a small sample). 
%Furthermore, a variety of phenomena of interest can be linked to such discrepancies, 
Other phenomena linked to such discrepancies include pluralistic ignorance, where individuals privately reject a view but believe the majority of other individuals accept it \cite{allport1924_socialpsyc_book}, the ``spiral of silence'' \cite{noelle1993spiral,taylor1982pluralistic_sos}, and enforcement of unpopular social norms \cite{centola2005selfenforce_norms,willer2009false_enforcement}. Whether occurring in a jury panel, a company boardroom or in the general population for a sensitive political issue, the potential societal ramifications of large and persistent discrepancies in private and expressed opinions are clear, and serve as a key motivator for our investigations.

%\subsection{Conformity: Empirical Data and Static Models}\label{ssec:static_intro}

\subsection{Existing Work}\label{ssec:existing_work}

\textit{Conformity: Empirical Data and Static Models.} One common reason such discrepancies arise is a pressure on an individual to conform in a group situation; formal study of such phenomena goes back over six decades. In 1951, Solomon E. Asch's seminal paper \cite{asch1951group_pressure_effects} showed an individual's public support for an indisputable fact could be distorted due to the pressure to conform to a unanimous group of others opposing this fact. 
%Other individuals publicly conceded to the group, but privately supported the fact, while yet others were uninfluenced by the unanimous group and maintained public support for the fact.
Asch's work was among the many studies examining the effects of pressures to conform to the group standard or opinion, using both controlled laboratory experiments and data gathered from field studies. Many of the lab experiments focus on Asch-like studies, perhaps with various modifications. A meta-analysis of 125 such studies was presented in \cite{bond2005group}. 
%Other studies found that high productivity factory workers can be pressured to lower their production rates to conform with factory averages \cite{coch1948overcoming}. 
Pluralistic ignorance is often associated with pressures to conform to social norms \cite{allport1924_socialpsyc_book,prentice1993pluralistic_alcohol,ogorman1975pluralistic}.
%, occurred among university students concerning their attitudes to alcohol drinking culture on campus \cite{prentice1993pluralistic_alcohol}, and among white Americans, concerning their attitudes to racial segregation in the 1960s \cite{ogorman1975pluralistic}. 
%while 	
%Driven by the empirical data, theoretical models of conformity were developed to explain and model the observations. 
With a focus on the seminal Asch experiments, a number of \emph{static} models were proposed to describe a single individual conforming to a unanimous majority \cite{tanford1984social_asch_model,mullen1983asch_model,stasser1981group_asch_model}, with obvious common limitations in generalisation to dynamics on social networks. 
%	They cannot be generalised to arbitrary social networks in which individuals interact only with neighbours as opposed to every other person in the group, despite network topology having been identified as playing an important role in how opinions and norms spread over larger groups \cite{kempe2003_maxspread}. Importantly, these \emph{static} models are \emph{unable to capture temporal aspects of conformity}, which can be significant \cite{noelle1993spiral,taylor1982pluralistic_sos,prentice1996pluralistic_chapter}. }
%The ``spiral of silence'' \cite{noelle1993spiral,taylor1982pluralistic_sos} focuses on how individuals may become less willing to express opinions \emph{due to the majority's views shifting over time}, while Prentice and Miller  \cite{prentice1996pluralistic_chapter} observed that ``many consequences of pluralistic ignorance occur \emph{over time} ... lead individuals to change their personal beliefs and opinions''. 

%\subsection{Agent-Based Opinion Dynamics Models}\label{ssec:abm_intro}

\textit{Opinion Dynamics Models.} Agent-based models (ABMs) have proved to be both versatile and powerful, with simple agent-level dynamics leading to interesting emergent network-level social phenomena.
%, and their analysis allows the identification and explanation of the network properties and socio-psychological processes that give rise to different phenomena over a network. 
The seminal French--DeGroot model \cite{french1956_socialpower,degroot1974OpinionDynamics} showed that a network of individuals can reach a consensus of opinions via weighted averaging of their opinions, a mechanism modelling ``social influence''. Indeed, the term ``influence network'' arose to reflect the social influence exerted via the interpersonal network. Since then, the roles of homophily \cite{hegselmann2002opinion,su2017HK_noise}, bias assimilation \cite{dandekar2013biased_degroot}, social distancing \cite{mas2014cultural}, and antagonistic interactions \cite{altafini2013antagonistic_interactions,proskurnikov2016opinion} in generating clustering, polarisation, and disagreement of opinions in the social network have also been studied. 
	%The impact of influence networks on the phenomenon of ``the wisdom of crowds'' has been reported in \cite{becker2017_crowdwisdom}. The effects of an individual's desire to be unique were studied in \cite{mas2014cultural,smaldino2015opdyn_conform_distinct}. 
	Individuals who remain somewhat attached to their initial opinions were introduced in the Friedkin--Johnsen model \cite{friedkin1990_FJsocialmodel} to explain the persistent disagreements observed in real communities. 
	%We note that the Friedkin--Johnsen model has extensive experimental validation, see e.g.  \cite{friedkin2011social_book,friedkin2017_truthwins_pnas,friedkin2016tevo_power}.  
	However, a key assumption in most existing ABMs (including those above), is that \textit{each individual has a single opinion} for a given topic. These models are unable to capture phenomena in which an individual holds, for the same topic, a private opinion different to the opinion he or she expresses.
	
	A few complex ABMs do exist in which each agent has both an expressed opinion and a private opinion for the same topic. The work \cite{centola2005selfenforce_norms} studies norm enforcement and assumes that each agent has two binary variables representing private and public acceptance or rejection of a norm. We are motivated to consider opinions as continuous variables to better capture \emph{discrepancies} in expressed and private opinions, since an individual's opinion may range in its intensity. The model in \cite{duggins2017_psych_opdyn} does assume the expressed and private opinions take values in a continuous interval, but is extremely complex and nonlinear. The properties of the models in \cite{centola2005selfenforce_norms,duggins2017_psych_opdyn} have only been partially characterised by simulation-based analysis, which is computationally expensive if detailed analysis is desired. 
	
		We seek to expand from \cite{centola2005selfenforce_norms,duggins2017_psych_opdyn} to build an ABM of lower complexity that is still powerful enough to capture how discrepancies in expressed and private opinions might evolve in social networks, and to allow study by theoretical analysis, as opposed to only by simulation. 
%		ABMs that are of relatively low complexity, but still accurate enough to capture relevant social phenomena, are desirable for several obvious reasons. 
%		Such models can be studied theoretically, as opposed to only by simulation. Often, and as we will show in this paper, insightful and illuminating social interpretations are secured.
		 Importantly also, a minimal number of parameters per agent makes data fitting and parameter estimation in experimental investigations a tractable process, as highlighted by the successful validations of the Friedkin--Johnsen model \cite{friedkin2011social_book,friedkin2017_truthwins_pnas,friedkin2016tevo_power}, whereas experiments for more complicated models are rare. 
	
	%ABMs that are of low complexity, but still accurate enough to capture relevant social phenomena, are desirable for several reasons. Such models can be theoretically analysed using systems and control theory, leading to the drawing of semi-quantitative conclusions on how key parameters such as the network topology affect the evolution of opinions. Often, insightful and illuminating social interpretations are secured. Importantly, low complexity typically implies a minimal number of parameters per agent, which makes data fitting and parameter estimation in experimental validations a tractable process. This is highlighted by the numerous successful validations of the Friedkin--Johnsen model \cite{friedkin2011social_book,friedkin2017_truthwins_pnas,friedkin2016tevo_power}, whereas experiments for more complicated models are rare. We are thus motivated to go beyond the works of \cite{centola2005selfenforce_norms} and \cite{duggins2017_psych_opdyn} to build an ABM of low complexity that is still powerful enough to study how discrepancies in expressed and private opinions might evolve in social networks. 

\subsection{Contributions of This Paper}\label{ssec:novel_intro}
In this paper, we aim to bridge the gap between the literature on conformity and the opinion dynamics models, by proposing a model where each individual (agent) has both a private and an expressed opinion. Inspired by the Friedkin--Johnsen model, we propose that an individual's private opinion evolves under social influence exerted by the individual's network neighbours' expressed opinions, but each individual remains attached to his or her initial opinion with a level of stubbornness. Then, and motivated by existing works on the pressures to conform in a group situation, we propose that each individual has some resilience to this pressure, and each individual expresses an opinion \emph{altered} from his or her private opinion to be closer to the average expressed opinion.
%Our model simultaneously advances on the traditional works of both studies of conformity in groups, and agent-based models of social influence.

Rigorous analysis of the model is given, leading to a number of semi-quantitative conclusions with insightful social interpretations. We show that for strongly connected networks and almost all parameter values for stubbornness and resilience, individuals' opinions converge exponentially fast to a steady-state of persistent disagreement.
% that depends on the individuals' initial private opinions, but not their initial expressed opinions. 
We identify that the combination of (i) stubbornness, (ii) resilience, and (iii) connectivity of the network generically leads to every individual having a discrepancy between his or her limiting expressed and private opinions. We give a method for underbounding the disagreement among the limiting private opinions given limited knowledge of the network, and show that a change in an individual's resilience to the pressure has a propagating effect on every other individual's expressed opinion. 
%These results build a deeper and richer understanding of the key factors determining the behaviour of the network, beyond works such as \cite{centola2005selfenforce_norms,duggins2017_psych_opdyn} that use simulation. 
Last, we apply our model to the seminal experiments on conformity by Asch \cite{asch1951group_pressure_effects}. Asch recorded 3 different types of responses among test individuals who must choose between expressing support for an indisputable fact and siding with a unanimous majority claiming the fact to be false. We identify stubbornness and resilience parameter ranges for all 3 responses; this capturing of all 3 responses is a first among ABMs, and underlines our model's strength as a general framework for studying the evolution of expressed and private opinions.

Our work extends from (i) the static models of conformity, by generalising to opinion dynamics on arbitrary networks, and (ii) the dynamic agent-based models, by introducing mechanisms inspired by socio-psychological literature to model the expressed and private opinions of each individual separately. The result is a general modelling framework, which is shown to be consistent with empirical data, and may be used to further the study of phenomena involving discrepancies in private and expressed opinions in social networks.

The rest of the paper is structured as follows. The model is presented in Section~\ref{sec:model}, with theoretical results detailed in Section~\ref{sec:analysis_main}. Section~\ref{sec:asch} applies the model to Asch's experiments, with concluding remarks given in Section~\ref{sec:con}.

\section{A Novel Model of Opinion Evolution Under Pressure to Conform}\label{sec:model}

Before introducing the model, we define some notation, and introduce graphs, which are used to model the network of interpersonal influence.

%\subsection{Notations}\label{ssec:notation}
\emph{Notations:} The $n$-column vector of all ones and zeros is given by $\vect{1}_n$ and $\vect{0}_n$ respectively. The $n\times n$ identity matrix is given by $\mat{I}_n$. For a matrix $\mat{A} \in \mathbb{R}^{n\times m}$ (respectively a vector $\vect{a}\in\mathbb{R}^n$), we denote the $(i,j)^{th}$ element as $a_{ij}$ (respectively the $i^{th}$ element as $a_i$). 
%The transpose of $\mat{A}$ is denoted by $\mat{A}^\top$.
A matrix $\mat{A}$ is said to be nonnegative, denoted by $\mat{A} \geq 0$ (respectively positive, denoted by $\mat{A} > 0$) if all of its entries $a_{ij}$ are nonnegative (respectively positive). A nonnegative matrix $\mat{A}$ is said to be row-stochastic (respectively row-substochastic) if for all $i$, there holds $\sum_{j=1}^n a_{ij} = 1$ (respectively $\sum_{j=1}^n a_{ij} \leq 1$ and $\exists k : \sum_{j=1}^n a_{kj} <1$). 
%For two matrices $\mat{A,B}\geq 0$, we say $\mat{A}\in \mathbb{R}^{n\times m}$ and $\mat{B}\in \mathbb{R}^{n\times m}$ are of the same type, denoted by $\mat{A} \sim \mat{B}$, if $\mat{A}$ and $\mat{B}$ have strictly positive elements at the same positions. 

%\subsection{Graph Theory}\label{ssec:graph}
\emph{Graphs:} Given any nonnegative not necessarily symmetric $\mat{A} \in \mathbb{R}^{n\times n}$, we can associate with it a graph $\mathcal{G}[\mat{A}] = (\mathcal{V}, \mathcal{E}[\mat{A}], \mat{A})$. Here, $\mathcal{V} = \{v_1, \hdots, v_n\}$ is the set of nodes, with index set $\mathcal{I} = \{1, \hdots, n\}$. An edge $e_{ij} = (v_i, v_j)$ is in the set of ordered edges $\mathcal{E}[\mat{A}] \subseteq \mathcal{V}\times \mathcal{V}$ if and only if $a_{ji} > 0$. The edge $e_{ij}$ is said to be incoming with respect to $j$ and outgoing with respect to $i$. 
%Because $\mat{A}$ is not assumed to be symmetric, in general, existence of $e_{ij}$ does not imply existence of $e_{ji}$. 
We allow self-loops, i.e. $e_{ii}$ is allowed to be in $\mathcal{E}$. The neighbour set of $v_i$ is denoted by $\mathcal{N}_i = \{v_j \in \mathcal{V} : (v_j,v_i) \in \mathcal{E} \}$. A directed path is a sequence of edges of the form $(v_{p_1}, v_{p_2}), (v_{p_2}, v_{p_3}), ...,$ where $v_{p_i} \in \mathcal{V}, e_{p_j p_k} \in \mathcal{E}$. 
%Node $i$ is reachable from node $j$ if there exists a directed path from $v_j$ to $v_i$. 
A graph $\mathcal{G}[\mat{A}]$ is strongly connected if and only if there is a path from every node to every other node \cite{godsil2001algebraic}, or equivalently, if and only if $\mat{A}$ is irreducible \cite{godsil2001algebraic}. A cycle is a directed path that starts and ends at the same vertex, and contains no repeated vertex except the initial (also the final) vertex, and a directed graph is \emph{aperiodic} if there exists no integer $k > 1$ that divides the length of every cycle of the graph \cite{bullo2009distributed}. 

We are now ready to propose the agent-based model. For a population of $n$ individuals, let $y_i(t) \in \mathbb{R}$ and $\hat{y}_i(t)\in \mathbb{R}$, $i = 1, \hdots, n$, represent, at time $t = 0, 1, \hdots$, individual $i$'s private and expressed opinions on a given topic, respectively. In general, $y_i(t)$ and $\hat{y}_i(t)$ are not the same, and \emph{we regard $y_i$ as individual $i$'s true opinion.} Individual $i$ may refrain from expressing $y_i(t)$ for many reasons, e.g. political correctness when discussing a sensitive topic. For instance, \emph{preference falsification} \cite{kuran1997pref_falsification} occurs when an individual falsifies his or her view due to social pressure (be it imaginary or real), or deliberately, e.g. by a politician seeking to garner votes. In our model, an individual falsifies his or her opinion due to a pressure to conform to the group average opinion. 
%As is common in agent-based opinion models \cite{degroot1974OpinionDynamics,friedkin2011social_book} we assume that the opinions take continuous values (i.e. in $\mathbb{R}$). 
The terms ``opinion'', ``belief'', and ``attitude'' all appear in the literature, with various related definitions
%\footnote{Different definitions may arise because the topic might be classified to be an intellective statement i.e. provably true or false, or a subjective question, i.e. for which no exact answer exists. 
%	An intellective topic might be ``smoking tobacco damages your lungs'' while `` was the 2003 US-led invasion of Iraq justified \cite{friedkin2016network_science}?'' would be a subjective question.}. 
Our model is general enough to cover all these terms, since in all such instances, one can scale $y_i(t), \hat{y}_i(t)$ to be in some real interval $[a,b]$, where $a$ and $b$ represent the two extreme positions on the topic. For consistency, we will \emph{only use ``opinion''} unless explicitly stated otherwise.
	
\begin{figure}
	\centering
	\includegraphics[width=0.65\linewidth]{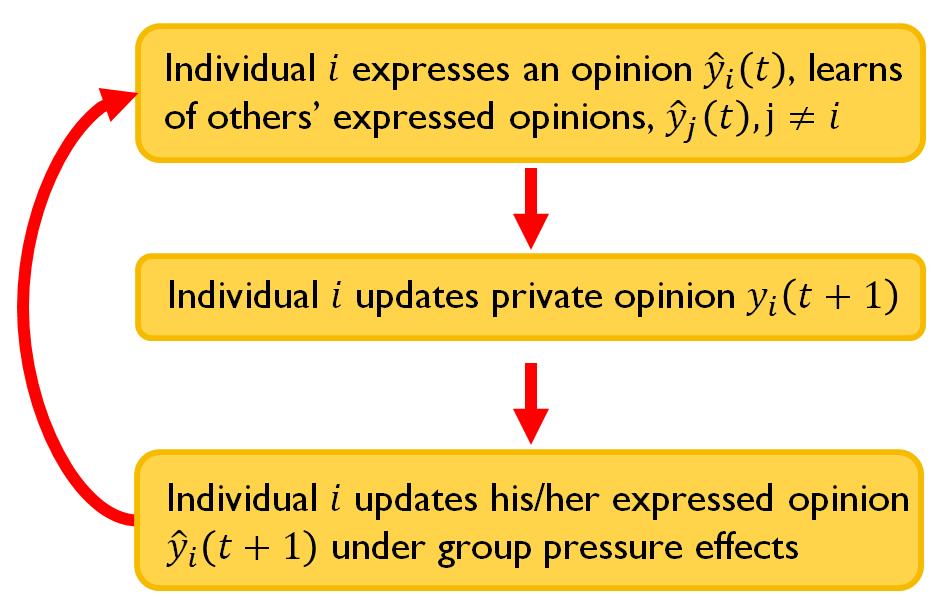}
	\caption{The discussion process. Each individual $i$, at time step $t$, expresses opinion $\hat{y}_i(t)$ and learns of others' expressed opinions $\hat{y}_j(t), j\neq i$. Next, the privately held opinion $y_i(t+1)$ evolves according to \eqref{eq:private_op}. After this, individual $i$ then determines the new $\hat{y}_i(t+1)$ to be expressed in the next round of discussion, according to \eqref{eq:public_op}.  }
	\label{fig:flow_chart}
\end{figure}

The individuals discuss their expressed opinions $\hat{y}_i(t)$ over a network described by a graph $\mathcal{G}[\mat{W}]$, and as a result, their private and expressed opinions, $y_i(t)$ and $\hat{y}_i(t)$ evolve in a process qualitatively described in Fig.~\ref{fig:flow_chart}.
%, and formally defined in \eqref{eq:private_op} and \eqref{eq:public_op} below.  
%The model for the evolution of $y_i(t)$ is inspired by the Friedkin--Johnsen model, while the model for how individual $i$ determines expressed opinion $\hat{y}_i(t)$ is proposed to reflect observations in the literature \cite{asch1951group_pressure_effects,gorden1952interaction_pressure} of how individuals deal with a pressure to conform to a group norm or standard. 
Formally, individual $i$'s private opinion evolves as 
\begin{equation}\label{eq:private_op}
y_i(t+1) = \lambda_i w_{ii}y_i(t) + \lambda_i \sum_{j\neq i}^n w_{ij} \hat{y}_j(t) + (1 - \lambda_i)y_i(0)
\end{equation}
and expressed opinion $\hat{y}_i(t)$ is determined according to
\begin{equation}\label{eq:public_op}
\hat{y}_i(t) = \phi_i y_i(t) + (1-\phi_i)\hat{y}_{i,\text{lavg}}(t-1).
\end{equation} 
In \eqref{eq:private_op}, the influence weight that individual $i$ accords to individual $j$'s expressed opinion $\hat{y}_j(t)$ is captured by $w_{ij} \geq 0$, satisfying $\sum_{j=1}^n w_{ij} = 1$ for all $i \in \mathcal{I}$. The term $w_{ii} \geq 0$ represents the self-confidence (if any) of individual $i$ in $i$'s own private opinion\footnote{In most situations, one can assume $w_{ii} > 0$, and models for studying the dynamics of $w_{ii}$ exist \cite{jia2015opinion_SIAM,ye2019DF_journal}. Presence of $w_{ii} > 0$ can also ensure convergence of the opinions, e.g. in the DeGroot model \cite{proskurnikov2017tutorial}.}.
%In particular, if $w_{ij} > 0$, then there is an edge $(v_j, v_i)$ in $\mathcal{G}[\mat{W}]$, and the edge connotes that individual $i$ learns of, and is influenced by the expressed opinion $\hat{y}_j$ of individual $j$. If $w_{ii} > 0$, then individual $i$ is influenced by $i$'s own private opinion. 
%We assume that for any $i$, $\sum_{j=1}^n w_{ij} = 1$. 
The constant $\lambda_i \in [0,1]$ represents individual $i$'s \emph{susceptibility to interpersonal influence} changing $i$'s private opinion ($1-\lambda_i$ is thus $i$'s stubbornness regarding initial opinion $y_i(0)$). Individual $i$ is maximally or minimally susceptible if $\lambda_i = 1$ or $\lambda_i = 0$, respectively. In \eqref{eq:public_op}, the quantity $\hat{y}_{i,\text{lavg}}(t) = \sum_{j\in \mathcal{N}_i} m_{ij}\hat{y}_i(t)$ is specific to individual $i$, and includes only the expressed $\hat{y}_j(t)$ of $i$'s neighbours. We assume that the weight $m_{ij} \geq 0$ satisfies $w_{ij} > 0 \Leftrightarrow m_{ij} > 0$ and $\sum_{j\in \mathcal{N}_i} m_{ij} = 1$; the matrix $\mat M = \{m_{ij}\}$ is therefore row-stochastic and $\mathcal{G}[\mat{M}]$ has the same connectivity properties as $\mathcal{G}[\mat{W}]$. A natural choice is $m_{ij} = \vert \mathcal{N}_i\vert^{-1}$ for all $j : e_{ji} \in \mathcal{E}[\mat W]$, while a reasonable alternative is $m_{ij} = w_{ij}, \forall i,j \in \mathcal{I}$. Thus, $\hat{y}_{i,\text{lavg}}(t)$ represents the group standard or norm as viewed by individual $i$ at time $t$, and is termed the \emph{local public opinion} as perceived by individual $i$. 
%	Where there is no confusion, we refer to $\hat{y}_{i,\text{lavg}}(t)$ simply as the \textit{public opinion}. 
The constant $\phi_i \in [0,1]$ encodes individual $i$'s \emph{resilience to pressures to conform} to the local public opinion (maximally 1, and minimally 0), or \emph{resilience} for short. The initial expressed opinion is set to be $\hat{y}_i(0) = y_i(0)$, which means \eqref{eq:private_op} comes into effect for $t=1$. As it turns out, under mild assumptions on $\lambda_i$, the final opinion values are dependent on $y_i(0)$ but independent of $\hat{y}_i(0)$; one could also select other initialisations for $\hat{y}_i(0)$ with the final opinions unchanged (though the transient would change).  

%Because $\sum_{j=1}^n w_{ij} = 1$ and $\lambda_i \in [0,1]$, \eqref{eq:private_op} indicates that $y_i(t+1)$ is a convex combination of $1)$ individual $i$'s initial private opinion $y_i(0)$, $2)$ $i$'s current private opinion $y_i(t)$, and $3)$ the expressed opinions of $i$'s neighbours, $\hat{y}_j(t)$. Similarly, $\hat{y}_i(t)$ is a convex combination of $y_i(t)$ and the public opinion at the previous time step, $\hat{y}_{\text{avg}}(t-1)$. 
%The influences that act to change individual $i$'s private and expressed opinions are shown in  Fig.~\ref{fig:individual_process}.  

Sociology literature indicates that the pressure to conform causes an individual to express an opinion that is in the direction of the perceived group standard \cite{asch1951group_pressure_effects,gorden1952interaction_pressure,taylor1982pluralistic_sos}, which in our model is $\hat{y}_{i,\text{lavg}}(t)$. Some pressures of conformity may derive from unspoken traditions \cite{merei1949group_leadership}, or a fear or being different \cite{asch1951group_pressure_effects}, and others arise because of a desire to be in the group, driven by e.g. monetary incentives, status or rewards \cite{festinger1950informal}. 
%At other times punishment is dealt to individuals who deviate from group norms to force conformity \cite{thrasher1927gang}, even if the norm itself is destructive to the group  \cite{abbink2017_badsocialnorms}. 
Thus, \eqref{eq:public_op} aims to capture individual $i$ expressing an opinion equal to $i$'s private opinion \emph{modified or altered due to normative pressure} (proportional to $1-\phi_i$) to be closer to the public opinion as perceived by individual $i$, which exerts a ``force'' $(1-\phi_i)\hat{y}_{i,\text{lavg}}(t-1)$. Heterogeneous $\phi_i$ captures the fact that some individuals are less inhibited/reserved than others when expressing their opinions. In addition, pressures are exerted (or perceived to be exerted), differentially for individuals, e.g. due to status \cite{schachter1951deviation,gorden1952interaction_pressure}. 

\begin{remark}\label{rem:global}
	 Use of a local public opinion $\hat{y}_{i,\text{lavg}}(t)$ ensures the model's scalability to large networks, but in small networks, e.g. a boardroom of 10 people, one could replace $\hat{y}_{i,\text{lavg}}(t)$ with the global public opinion $\hat{y}_{\text{avg}}(t) = \frac{1}{n} \sum_{j=1}^n \hat{y}_j(t)$ since it is likely to be discernible to every individual. It turns out that all but one of the high-level theoretical conclusions, including convergence, do not depend on the choice of weights of the local public opinion, nor on whether a local or global public opinion is used. However, preliminary observations show that the distribution of the final opinion values can vary significantly depending on the aforementioned choices, and we leave characterisation of the difference to future investigations.
\end{remark}

%The motivation for modelling stubbornness in this manner is discussed in e.g. \cite{friedkin2011social_book}. 

\begin{remark}
A key feature in our model, departing from most existing models, is the associating of two states $y_i, \hat{y}_i$ for each individual and the restriction that only other $\hat{y}_j$ (and no $y_j$) may be available to individual $i$. Importantly, note that $\hat y_i(t)$ evolves \textit{dynamically} via \eqref{eq:public_op}; $\hat y_i(t)$ is not simply an output variable.
%This paper therefore departs from many established models, which assume only one opinion variable per individual (we covered a variety of works in the introduction). 
However, notice that setting $\phi_i = 1$ for all $i$ recovers the Friedkin--Johnsen model, while $\phi_i = \lambda_i = 1$ for all $i$, recovers the DeGroot model \cite{degroot1974OpinionDynamics}. One may also notice the time-shift in \eqref{eq:public_op} of $\hat{y}_{i, \text{lavg}}(t-1)$, which ensures that \eqref{eq:public_op} is consistent with the qualitative process described in Fig.~\ref{fig:flow_chart}. Thus, \eqref{eq:public_op} aims to capture a natural manner, widely supported in the sociology literature, in which an individual determines his or her expressed opinion under a pressure to conform. 
%Last, the use of $\hat{y}_{i, \text{lavg}}(t-1)$ means that $\hat y_i(t)$ is uniquely defined for all $i$ when $\phi_i = 0, \forall\, i$, whereas if $\hat{y}_{i, \text{lavg}}(t)$ were used, this is not the case. 
\end{remark}

%\begin{figure}
%	\centering
%	\resizebox{0.875\linewidth}{!}{\input{individual_process.pdf_tex}}
%	\caption{Individual $i$'s private opinion $y_i(t)$ and expressed opinion $\hat{y}_i(t)$ are represented by the blue node and purple node, respectively. The edges represent an influence exerted on $i$'s private or expressed opinion. In particular, $y_i(t)$ is changed due to influence from others' expressed opinion $\hat{y}_j(t)$ individuals in the influence network, $i$'s self-weight (if there is a self-loop) and an attachment to initial opinion $y_i(0)$. The private opinion $y_i(t)$ exerts an influence on $i$'s expressed opinion $\hat{y}_i(t)$ via a resilience, while the public opinion $\hat{y}_{\text{avg}}(t-1)$ exerts a pressure to conform. Individual $i$ exerts an influence on others in the network via $i$'s expressed opinion $\hat{y}_i$. }
%	\label{fig:individual_process}
%\end{figure}

%Later, we identify using theoretical analysis that differences in the opinion distributions of the network at equilibrium can arise depending on whether \eqref{eq:public_op_main} or \eqref{eq:public_op_local_main} is used to model the expressed opinion. These differences can be significant, as discovered in Section~\ref{sec:pluralistic}, when we use extensive simulations to investigate the role of stubborn extremists in creating pluralistic ignorance.

\subsection{The Networked System Dynamics}
We now obtain a matrix form equation for the dynamics of all individuals' opinions on the network. Let $\vect{y} = [y_1, y_2, \hdots, y_n]^\top$ and $\hat{\vect{y}} = [\hat{y}_1, \hat{y}_2, \hdots, \hat{y}_n]^\top$ be the stacked vectors of private and expressed opinions $y_i$ and $\hat{y}_i$ of the $n$ individuals in the influence network, respectively. The influence matrix $\mat{W}$ can be decomposed as $\mat{W} = \wt{\mat{W}} + \wh{\mat{W}}$ where $\wt{\mat{W}}$ is a diagonal matrix with diagonal entries $\tilde{w}_{ii} = w_{ii}$. The matrix $\wh{\mat{W}}$ has entries $\wh{w}_{ij} = w_{ij}$ for all $j\neq i$ and $\wh{w}_{ii} = 0$ for all $i$. Define $\mat{\Lambda} = diag(\lambda_i)$ and $\mat{\Phi} = diag(\phi_i)$. 
Substituting $\hat{y}_j(t)$ from \eqref{eq:public_op} into \eqref{eq:private_op}, and recalling that $\hat{y}_{i, \text{lavg}} = \sum_{j \in \mathcal{N}_j} m_{ij}\hat{y}_{j}$, yields
\begin{align}\label{eq:private_op_sub}
y_i(t+1) & \!=\! \lambda_i w_{ii}y_i(t) + \lambda_i  \sum_{j\neq i}^n w_{ij}\phi_j y_j(t) \!+ (1 - \lambda_i)y_i(0) \nonumber \\
& \quad + \lambda_i\sum_{j\neq i}^n w_{ij} (1-\phi_j) \sum_{k \in \mathcal{N}_j} m_{jk}\hat{y}_{k}(t-1). 
\end{align}
From \eqref{eq:private_op_sub} and \eqref{eq:public_op}, one obtains
\begin{align}\label{eq:system_compact}
\begin{bmatrix} \vect{y}(t+1) \\ \hat{\vect{y}}(t) \end{bmatrix} =
\mat{P}
\begin{bmatrix}
\vect{y}(t) \\ \hat{\vect{y}}(t-1)
\end{bmatrix}
+
\begin{bmatrix}
\left(\mat{I}_n - \mat{\Lambda}\right)\vect{y}(0) \\ \vect{0}_n
\end{bmatrix},
\end{align}
where $\mat{P}$ consists of the following block matrices
\begin{align}\label{eq:P_matrix}
\begin{bmatrix}
\mat{\Lambda}( \wt{\mat{W}} + \wh{\mat{W}}\mat{\Phi}) \hspace{3pt}& \mat{\Lambda}\wh{\mat{W}}(\mat{I}_n - \mat{\Phi} )\mat M \\
\mat{\Phi} & \left( \mat{I}_n - \mat{\Phi}\right) \mat M
\end{bmatrix} = \begin{bmatrix} \mat{P}_{11} & \mat{P}_{12} \\ \mat{P}_{21} & \mat{P}_{22} \end{bmatrix}
\end{align}
As stated above, we set the initialisation as $\hat{\vect y}(0) = \vect y(0)$, yielding $\vect y(1) = (\mat{\Lambda W} +\mat{I}_n -\mat \Lambda) \vect y(0)$.

\section{Analysis of the Opinion Dynamical System}\label{sec:analysis_main}
We now investigate the evolution of $y_i(t)$ and $\hat{y}_i(t)$, according to \eqref{eq:private_op} and \eqref{eq:public_op}, for the $n$ individuals interacting on the influence network $\mathcal{G}[\mat{W}]$. In order to place the focus on social interpretations, we first present the theoretical statements, and then discuss conclusions. All the proofs are deferred to the Appendix, since the key focus of this section is to secure conclusions via analysis of \eqref{eq:system_compact} regarding the \emph{discrepancies between expressed and private opinions} that form over time. Throughout this section, we make the following assumption on the social network. 

\begin{assumption}\label{assm:PE_network}
The network $\mathcal{G}[\mat{W}]$ is strongly connected and aperiodic, and $\mat{W}$ is row-stochastic. Furthermore, there holds $\lambda_i, \phi_i \in (0,1),\forall\,i\in\mathcal{I}$.
\end{assumption}
It should be noted that for the purpose of convergence analysis, almost certainly one could relax the assumption to include graphs which are not strongly connected, and for $\phi_i, \lambda_i \in [0,1]$, which we leave for future work.

Notice that because $\sum_{j=1}^n w_{ij} = 1$ and $\lambda_i \in [0,1]$, \eqref{eq:private_op} indicates that $y_i(t+1)$ is a convex combination of $y_i(0)$, $y_i(t)$, and $\hat{y}_j(t), j \in \mathcal{N}_i$. Similarly, $\hat{y}_i(t)$ is a convex combination of $y_i(t)$ and $\hat{y}_{i,\text{lavg}}(t-1)$. It follows that
\begin{equation}
\mathcal{S} = \{y_i, \hat{y}_i : \min_{k\in\mathcal{I}} y_k(0)\leq y_i, \hat{y}_i \leq \max_{j\in\mathcal{I}} y_j(0), i \in \mathcal{I}\}
\end{equation}
is a positive invariant set of the system \eqref{eq:system_compact}, which is a desirable property. If $y_i(0) \in [a,b]$, where $a,b \in \mathbb{R}$ represent the two extremes of the opinion spectrum, and $\mathcal{S}$ is a positive invariant set of \eqref{eq:system_compact}, then the opinions are always well defined.
%(it does not make sense for $y_i(t^\prime) < a$ or $y_i(t^\prime) > b$ for some $t^\prime \geq 0$ if $a,b$ are the upper and lower extremes).

%\begin{lemma}[Invariant Set]\label{lem:PE_nonexpansive_WRP_y0}
%	Consider a network $\mathcal{G}[\mat{W}]$ where each individual $i$'s opinions $y_i(t)$ and $\hat{y}_i(t)$ evolve according to \eqref{eq:private_op} and \eqref{eq:public_op}, respectively. Suppose Assumption~\ref{assm:PE_network} holds, and that $\vect{y}(0) = \hat{\vect{y}}(0)$.  Then, for all $t \geq 0$, there holds
%	\begin{align}
%	\max_{i, j\in \mathcal{I}} \left\{y_i(t), \hat{y}_j(t)\right\} & \le \max_{i\in\mathcal{I}} y_i(0) = \max_{i\in\mathcal{I}} \hat{y}_i(0), \label{eq:upbound_y}\\
%	\min_{i,j \in \mathcal{I}} \left\{y_i(t), \hat{y}_j(t) \right\}& \ge \min_{i\in \mathcal{I}} y_i(0) = \min_{i\in \mathcal{I}} \hat{y}_i(0). \label{eq:lowbound_y}
%	\end{align}
%\end{lemma}
%In the extended arXiv version \cite{ye2018_PublicPrivate_arXiv}, a simple simulation counter-example is given to show that \emph{there need not hold } $\max_{i, j\in \mathcal{I}} \left\{ y_i(t), \hat{y}_j(t)\right\} \geq \max_{i, j\in \mathcal{I}} \left\{y_i(t+1),\hat{y}_j(t+1)\right\} $, $\min_{i, j\in \mathcal{I}} \left\{ y_i(t), \hat{y}_j(t) \right\}  \leq \min_{i, j\in \mathcal{I}} \left\{ y_i(t+1), \hat{y}_j(t+1) \right\}$ for all $t \geq 0$. This is a semi-contractive property held by the DeGroot model, see e.g. \cite{proskurnikov2017tutorial}. 

\subsection{Convergence}
The main convergence theorem, and a subsequent corollary for consensus, are now presented.
\begin{theorem}[Exponential Convergence]\label{thm:PE_convergence_eqb}
	Consider a network $\mathcal{G}[\mat{W}]$ where each individual $i$'s opinions $y_i(t)$ and $\hat{y}_i(t)$ evolve according to \eqref{eq:private_op} and \eqref{eq:public_op}, respectively. Suppose Assumption~\ref{assm:PE_network} holds. Then, the system \eqref{eq:system_compact} converges exponentially fast to the limit
		\begin{align}
	\lim_{t\to\infty} \vect{y}(t) & \triangleq \vect{y}^*  = \mat{R}\vect{y}(0) \label{eq:final_private} \\
	\lim_{t\to\infty} \hat{\vect{y}}(t) & \triangleq \hat{\vect{y}}^*  = \mat{S}\vect{y}^*, \label{eq:final_expressed}
	\end{align}
	where $\mat{R} = (\mat{I}_n - (\mat{P}_{11} + \mat{P}_{12}\mat{S}))^{-1}(\mat{I}_n - \mat{\Lambda})$ and $\mat{S}  = (\mat{I}_n - \mat{P}_{22})^{-1}\mat{P}_{21}$ are positive and row-stochastic, with $\mat{P}_{ij}$ defined in \eqref{eq:P_matrix}. 
\end{theorem}

The above shows that the final private and expressed opinions depend on $\vect{y}(0)$, while $\hat{\vect y}(0)$ are forgotten exponentially fast; one could initialise $\hat{\vect y}(0)$ arbitrarily, though the transient will differ. The row-stochasticity of $\mat{R}$ and $\mat{S}$ implies that the final private and expressed opinions are a convex combination of the initial private opinions. Additionally, $\mat{R}, \mat S > 0$ means every individual $i$'s initial $y_i(0)$ has an influence on every individual $j$'s final opinions $y_j^*$ and $\hat{y}_j^*$, a reflection of the strongly connected network. The following corollary establishes a condition for consensus of opinions, though one notes that part of the hypothesis for Theorem~\ref{thm:PE_convergence_eqb} is discarded.

\begin{corollary}[Consensus of Opinions]\label{cor:PE_consensus}
	Suppose that $\phi_i \in (0,1)$, and $\lambda_i = 1$, for all $i\in \mathcal{I}$. Suppose further that $\mathcal{G}[\mat{W}]$ is strongly connected and aperiodic, and $\mat{W}$ is row-stochastic. Then, for the system \eqref{eq:system_compact}, $\lim_{t\to\infty} \vect{y}(t) = \lim_{t\to\infty} \hat{\vect{y}}(t) = \alpha \vect{1}_n$ for some $\alpha \in \mathbb{R}$, exponentially fast.
\end{corollary}

\subsection{Discrepancies and Persistent Disagreement}\label{ssec:difference_PEModel}
This section establishes how disagreement among the opinions at steady state may arise. 
%A key conclusion is that stubbornness and resilience create a discrepancy in the expressed and private opinions of any individual. 
In the following theorem, let $z_{\max} \triangleq \max_{i = 1, \hdots, n} z_i$ and $z_{\min} \triangleq \min_{i = 1, \hdots, n} z_i$ denote the largest and smallest element of $\vect{z}\in\mathbb{R}^n$.

\begin{theorem}\label{thm:PE_disagree}
	Suppose that the hypotheses in Theorem~\ref{thm:PE_convergence_eqb} hold. If $\vect{y}(0) \neq \alpha \vect{1}_n$ for some $\alpha \in \mathbb{R}$, then the final opinions obey the following inequalities
	\begin{subequations}\label{eq:final_ineq}
		\begin{align}
		y(0)_{\max} & >  y^*_{\max}  >  \hat{y}^*_{\max} \label{eq:max_ineq} \\
		y(0)_{\min} & < y^*_{\min}  < \hat{y}^*_{\min} \label{eq:min_ineq} 
		\end{align}
	\end{subequations}
	and $\hat{y}^*_{\min} \neq \hat{y}^*_{\max}$. Moreover, given a network $\mathcal{G}[\mat{W}]$ and parameter vectors $\vect{\phi} = [\phi_1, \hdots, \phi_n]^\top$ and $\vect{\lambda} = [\lambda_1, \hdots, \lambda_n]^\top$, the set of initial conditions $\vect{y}(0)$ for which precisely $m > 0$ individuals $i_j \in \{i_1, \hdots, i_m\}\subseteq \mathcal{I}$ have $y_{i_j}^* = \hat{y}_{i_j}^*$, i.e. $m \triangleq \vert \{i \in \mathcal{I} : y_i^* = \hat{y}_i^*\} \vert$, lies in a subspace of $\mathbb{R}^n$ with dimension $n-m$. 
\end{theorem}

This result shows that for generic initial conditions there is a persistent disagreement of final opinions at the steady-state. 
This is a consequence of individuals not being maximally susceptible to influence, $\lambda_i < 1\,\forall\,i\in\mathcal{I}$. 
%If on the other hand $\lambda_i = 1,,\forall\,i\in \mathcal{I}$ then a consensus of opinions is reached exponentially fast (see Corollary~\ref{cor:PE_consensus}). 
One of the key conclusions of this paper is that for \emph{any individual} $i$ in the network, $y_i^* \neq \hat{y}_i^*$ for generic initial conditions, which is a subtle but significant difference from \eqref{eq:final_ineq}. More precisely, \emph{the presence of both stubbornness and pressure to conform, and the strong connectedness of the network creates a discrepancy between the private and expressed opinions of an individual}.
%  in generic networks and with generic initial conditions (See Remark~\ref{rem:prob1} below for further comments)
Without stubbornness ($\lambda_i = 1,\forall\,i$), a consensus of opinions is reached, and without a pressure to conform ($\phi_i = 1$), an individual has the same private and expressed opinions. Without strong connectedness, some individuals will not be influenced to change opinions. 

One further consequence of \eqref{eq:final_ineq} is that $y_{\max}^* - y_{\min}^* > \hat{y}_{\max}^* - \hat{y}_{\min}^*$, which implies that the \emph{level of agreement is greater} among the final expressed opinions when compared to the final private opinions. In other words, individuals are more willing to agree with others when they are expressing their opinions in a social network due to a pressure to conform. Moreover, the extreme final expressed opinions are upper and lower bounded by the final private opinions, which are in turn upper and lower bounded by the extreme initial private opinions, showing the effects of interpersonal influence and a pressure to conform. 

\begin{remark}\label{rem:prob1}
	Theorem~\ref{thm:PE_disagree} states that generically, there will be no two individuals who have the same final private opinions, and no individual will have the same final private and expressed opinion. Let the parameters defining the system ($\mat{W}$, $\vect{\phi}$ and $\vect{\lambda}$) be given and suppose that one runs $p$ experiments with $y_i(0)$ sampled independently from a distribution (uniform, normal, beta, etc.) over a non-degenerate interval\footnote{A statistical distribution is degenerate if for some $k_0$ the cumulative distribution function $F(x,k_0) = 0$ if $x< k_0$ and $F(x,k_0) = 1$ if $x\geq k_0$.}. If $q$ is the number of those experiments which result in  $y_i^* = \hat{y}_i^*$ for some $i \in \mathcal{I}$, then $\lim_{p\to\infty} q/p = 0$. From yet another perspective, the set of $\vect{y}(0)$ for which $y_i^* = \hat{y}_i^*$ for some $i \in \mathcal{I}$ belongs in a subspace of $\mathbb{R}^n$ that has a Lebesgue measure of zero. Similarly, $y_i^* = y_j^*$ for $i\neq j$ generically.
\end{remark}

\subsection{Estimating Disagreement in the Private Opinions}\label{ssec:estimate_PEModel}

We now give a quantitative method for underbounding the disagreement in the steady-state private opinions for a special case of the model, where we replace the local public opinion $\hat y_{i, \text{lavg}}$ with the global public opinion $\hat y_{\text{avg}} = n^{-1}\sum_{j=1}^n \hat y_i$ in \eqref{eq:public_op} for all individuals. 

\begin{corollary}\label{cor:scrambling_constant}
	Suppose that, for all $i\in \mathcal{I}$, $\hat y_{i, \text{lavg}}(t-1)$ in \eqref{eq:public_op} is replaced with $\hat y_{\text{avg}} = n^{-1}\sum_{j=1}^n \hat y_i$. Let $\kappa(\vect{\phi}) = 1-\frac{\phi_{\min}}{\phi_{\max}}(1-\phi_{\max}) \in (0,1)$ and $\phi_{\max} = \max_{i\in \mathcal{I}} \phi_i$, $\phi_{\min} = \min_{i\in\mathcal{I}} \phi_i$. Suppose further that the hypotheses in Theorem~\ref{thm:PE_convergence_eqb} hold. Then,
	\begin{align}
	\frac{ \hat{y}^*_{\max} - \hat{y}^*_{\min} }{ \kappa(\vect{\phi})  } & \leq y^*_{\max} - y^*_{\min}. \label{eq:scrambling_bound}
	\end{align}
\end{corollary}

For the purposes of monitoring the level of unvoiced discontent in a network (e.g. to prevent drastic and unforeseen actions or violence \cite{goodwin2011arabspring,kuran1989revolutions,duggins2017_psych_opdyn}), it is of interest to obtain more knowledge about the level of disagreement among the private opinions: $y_{\max}^* - y_{\min}^*$. A fundamental issue is that such information is by definition unlikely to be obtainable (except in certain situations like the post-experimental interviews conducted by Asch in his experiments, see Section~\ref{sec:asch}).
On the other hand, one expects that the level of expressed disagreement $\hat{y}^*_{\max} - \hat{y}^*_{\min}$ may be available. While one cannot expect to know every $\phi_i$, we argue that $\phi_{\max}$ and $\phi_{\min}$ might be obtained, if not accurately then approximately. If the global public opinion $\hat y_{\text{avg}}$ acts on all individuals, then Corollary~\ref{cor:scrambling_constant} gives a method for computing a \emph{lower bound} on the level of private disagreement given some limited knowledge.
% of (i) the final expressed opinions, and (ii) an estimate of the resilience levels of the individuals. 

It is obvious that if $\kappa(\vect{\phi})$ is small (if $\phi_{\max}$ is small and the ratio $\phi_{\min}/\phi_{\max}$ is close to 1), then even strong agreement among the expressed opinions (a small $\hat{y}_{\max}^* - \hat{y}_{\min}^*$) does not preclude significant disagreement in the final private opinions of the individuals. This might occur in e.g., an authoritarian government. The tightness of the bound \eqref{eq:scrambling_bound} depends on the ratio $\phi_{\min}/\phi_{\max}$; the closer the ratio is to one (i.e. as the ``force'' of the pressure to conform felt by each individual becomes more uniform), the tighter the bound. 
%If $\phi_i\,\forall\,i$ are known, one can obtain $\vect y^* = \mat{S}^{-1}\hat{\vect y}^*$ precisely.

%If local public opinion updating is used, i.e. \eqref{eq:public_op_local}, then \eqref{eq:scrambling_bound} no longer holds; we conjecture that a similar result may be obtained, although with difficulty since $\mat{S}$ now also depends on $\mat{W}$, in addition to $\mat{\phi}$.

\subsection{An Individual's Resilience Affects Everyone}\label{ssec:resilience_PEModel}
An interesting result is now presented, that shows how individual $i$'s resilience $\phi_i$ is propagated through the network.
% to affect the final expressed opinions of all other individuals.

%\begin{corollary}[Individual Resilience]\label{cor:derivative}
%	Suppose that the hypotheses in Theorem~\ref{thm:PE_convergence_eqb} hold. Then, the matrix $\mat{S}\in \mathbb{R}^{n\times n}$ appearing in \eqref{eq:final_expressed} is a function of $\phi_i, i\in \mathcal{I}$ and has partial derivative $\frac{\partial(\mat{S})} {\partial \phi_i} \in \mathbb{R}^{n\times n}$ with the following sign pattern \vspace*{-11pt}
%	\begin{equation}\label{eq:derivative}
%	\frac{\partial(\mat{S})}{\partial \phi_i}  = \! \!\! \!
%	\bbordermatrix{ & ~ & ~ & ~ & \! i^{th} column \! & ~ & ~ & ~ \cr
%		& - & \hdots & - & + & - & \hdots & - \cr
%		%& - & \hdots & - & + & - & \hdots & - \cr 
%		& \vdots & \ddots & \vdots & \vdots & \vdots & \ddots & \vdots \cr 
%		%& - &  \hdots & - & + & - & \hdots & - \cr 
%		& - &  \hdots & - & + & - & \hdots & - \cr 
%	}
%	\end{equation}
%	That is, $\frac{\partial(\mat{S})} {\partial \phi_i}$ has positive entries in the $i^{th}$ column and all other entries are negative.
%\end{corollary}

\begin{corollary}\label{cor:derivative}
	Suppose that the hypotheses in Theorem~\ref{thm:PE_convergence_eqb} hold. Then, the matrix $\mat{S}$ in \eqref{eq:final_expressed} has partial derivative $\frac{\partial(\mat{S})} {\partial \phi_i}$ with strictly positive entries in the $i^{th}$ column and with all other entries strictly negative.
\end{corollary}

Recall below Theorem~\ref{thm:PE_convergence_eqb} that individual $k$'s final expressed opinion $\hat{y}_k^*$ is a convex combination of all individuals' final private opinions $y_j^*$, with convex weights $s_{kj}$, $j = 1, \hdots, n$. Intuitively, increasing $\phi_k$ makes individual $k$ more resilient to the pressure to conform, and this is confirmed by the above; $\frac{\partial s_{kk}}{\partial \phi_k} > 0$ and $\frac{\partial s_{kj}}{\partial \phi_k} < 0$ for any $j \neq k$ and thus $\hat{y}_k^* \to y_k^*$ as $\phi_k \to 1$. 

More importantly, the above result yields a surprising and nontrivial fact; \emph{every entry} of the $k^{\text{th}}$ column of  $\frac{\partial(\mat{S})} {\partial \phi_k}$ is strictly positive, and all other entries of  $\frac{\partial(\mat{S})} {\partial \phi_k}$ are strictly negative. In context, any change in individual $k$'s resilience directly impacts every other individual's final expressed opinion due to the network of interpersonal influences. In particular, as $\phi_k$ increases (decreases), an individual $j$'s final expressed opinion $\hat{y}_j^*$ becomes closer to (further from) the final private opinion $y_k^*$ of individual $k$, since  $\frac{\partial s_{jk}}{\partial \phi_k} > 0$ (decreasing, since $\frac{\partial s_{jk}}{\partial \phi_k} < 0$).

\subsection{Simulations}\label{ssec:simulations_PEModel}
Two simulations are now presented to illustrate the theoretical results. A $3$-regular network\footnote{A $k$-regular graph is one which every node $v_i$ has $k$ neighbours, i.e. $\vert \mathcal{N}_i \vert = k\,\forall\,i\in\mathcal{I}$.} $\mathcal{G}[\mat{W}]$ with $n=18$ is generated. Self-loops are added to each node (to ensure $\mathcal{G}[\mat{W}]$ is aperiodic), and the influence weights $w_{ij}$ are obtained as follows. The value of each $w_{ij}$ is drawn randomly from a uniform distribution in the interval $(0,1)$ if $(v_j,v_i) \in \mathcal{E}$, and once all $w_{ij}$ are determined, the weights are normalised by dividing all entries in row $i$ by $\sum_{j=1}^n w_{ij}$. This ensures that $\mat{W}$ is row-stochastic and nonnegative. For $i\neq j$, it is not required that $w_{ij} = w_{ji}$ (which would result in an undirected graph), but for simplicity and convenience the simulations impose\footnote{Such an assumption is not needed for the theoretical results, but is a simple way to ensure that all directed graphs generated using the MATLAB package are strongly connected.} that $w_{ij} > 0 \Leftrightarrow w_{ji} > 0$. The values of $y_i(0)$, $\phi_i$, and $\lambda_i$, are selected from beta distributions, which have two parameters $\alpha$ and $\beta$. For $\alpha, \beta > 1$, a beta distribution of the variable $x$ is unimodal and satisfies $x \in (0,1)$, which is precisely what is required to satisfy Assumption~\ref{assm:PE_network} regarding $\phi_i, \lambda_i$. The beta distribution parameters are (i) $\alpha = 2$, $\beta = 2$ for $y_i(0)$, (ii) $\alpha = 2$, $\beta = 2$ for $\phi_i$, and (iii) $\alpha = 2$, $\beta = 8$ for $\lambda_i$. In the simulation, we use the global public opinion model (see Remark~\ref{rem:global}) to also showcase Corollary~\ref{cor:scrambling_constant}.

\begin{figure}
	\centering
	\includegraphics[height=0.9\linewidth,angle=-90]{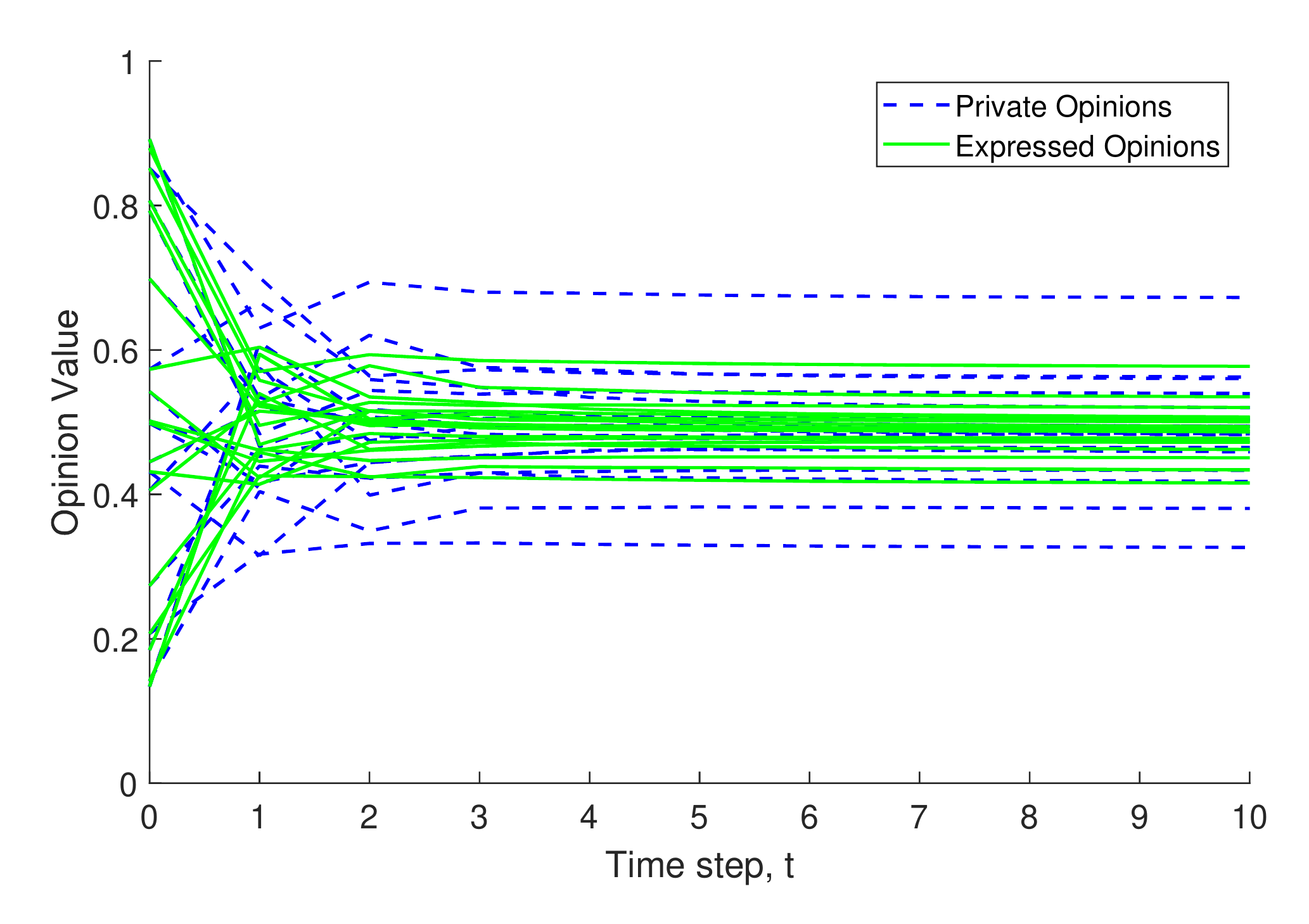}
	\caption{ Temporal evolution of opinions for 18 individuals  in an influence network. The green and dotted blue lines represent the expressed and private opinions of the individuals, respectively. }
	\label{fig:PE_model_example}
\end{figure}

The temporal evolution of opinions is shown in Fig.~\ref{fig:PE_model_example}. Several of the results detailed in this section can be observed. In particular, it is clear that \eqref{eq:final_expressed} holds. That is, there is no consensus of the limiting expressed or private opinions. Moreover, the disagreement among the final expressed opinions, $\hat{y}^*_{\max} - \hat{y}^*_{\min}$, is strictly smaller than the disagreement among the final private opinions, $y^*_{\max} - y^*_{\min}$. Separate to this, the final private opinions enclose the final expressed opinions from above and below. For the given simulation, the largest and smallest resilience values are $\phi_{\max} = 0.9437$ and $\phi_{\min} = 0.1994$, respectively. This implies that $\kappa(\vect{\phi}) = 0.9881$. One can also obtain that $\hat{y}^*_{\max} - \hat{y}^*_{\min} = 0.1613$. From \eqref{eq:scrambling_bound}, this indicates that  $y^*_{\max} - y^*_{\min} \geq 0.163$. The simulation result is consistent with the lower bound, in that $y^*_{\max} - y^*_{\min} = 0.3455$. Also, the bound is not tight, since $\phi_{\min}/\phi_{\max}$ is far from 1 (see Section~\ref{ssec:estimate_PEModel}).

\begin{figure}
	\centering
	\includegraphics[height=0.9\linewidth,angle=-90]{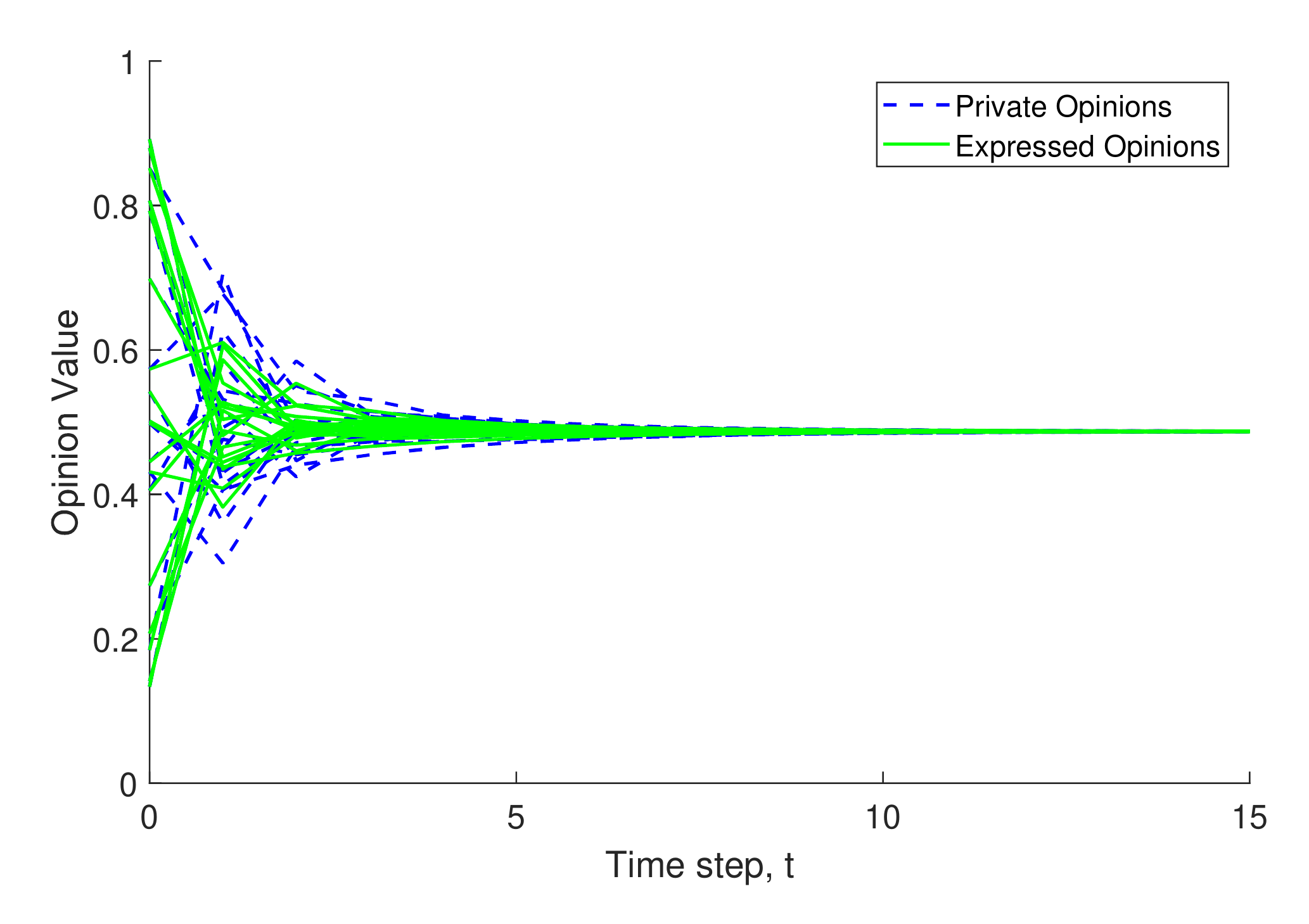}
	\caption{Temporal evolution of opinions for 18 individuals  in an influence network. The green and dotted blue lines represent the expressed and private opinions of the individuals, respectively. The lack of stubbornness,  $\lambda_i = 1,\forall\,i$, means that all opinions reach a consensus. }
	\label{fig:PE_model_example_consensus}
\end{figure}

For the same $\mathcal{G}[\mat{W}]$, with the same initial conditions $y_i(0)$ and resilience $\phi_i$, a second simulation is run with $\lambda_1 = 1,\forall i \in \mathcal{I}$. As shown in Fig.~\ref{fig:PE_model_example_consensus}, the opinions converge to a consensus $\vect{y}^* = \hat{\vect y}^* = \alpha \vect{1}_n$, for some $\alpha \in \mathbb{R}$, which illustrates Corollary~\ref{cor:PE_consensus}.

\section{Application to Asch's Experiments}\label{sec:asch}

We now use the model to revisit Solomon E. Asch's seminal experiments on conformity \cite{asch1951group_pressure_effects}. There are at least two objectives. For one, successfully capturing Asch's empirical data constitutes a form of soft validation for the model. Second, we aim to identify the values of the individual's susceptibility $\lambda_i$ and resilience $\phi_i$ that determine the individual's reaction to a unanimous majority's pressure to conform, and thus give an agent-based model explanation of the recorded observations. In order for the reader to fully appreciate and understand the results, a brief overview of the experiments and its results are now given, and the reader is referred to \cite{asch1951group_pressure_effects} for full details on the results. In summary, the experiments studied an individual's response to ``two contradictory and irreconcilable forces'' \cite{asch1951group_pressure_effects} of (i) a clear and indisputable fact, and (ii) a unanimous majority of the others who take positions opposing this fact. 

\begin{figure}
	\centering
	\includegraphics[width=0.35\linewidth]{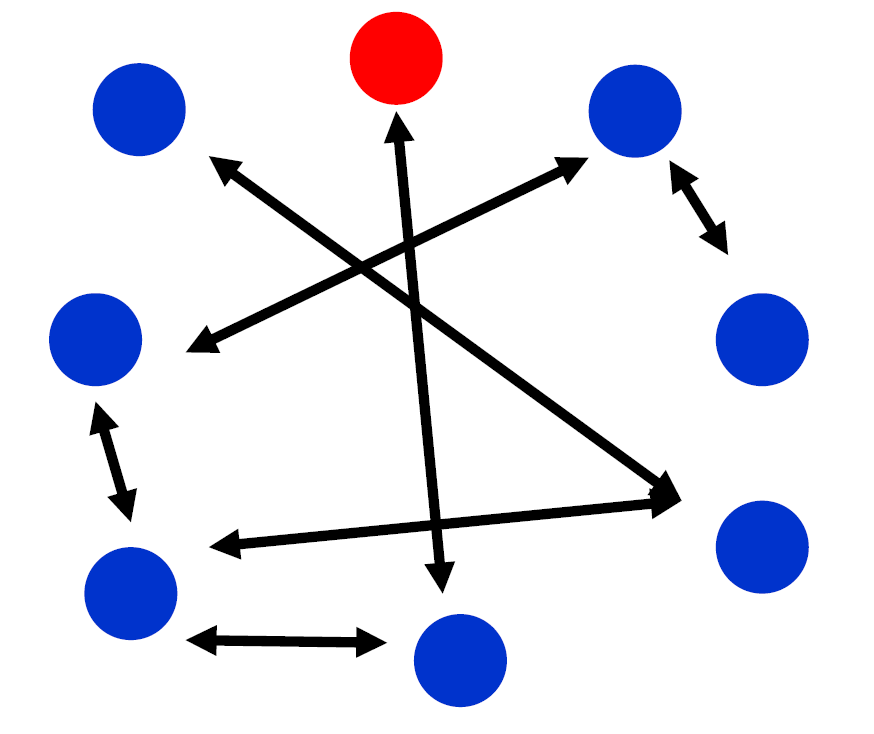} \hfill
	\includegraphics[width=0.6\linewidth]{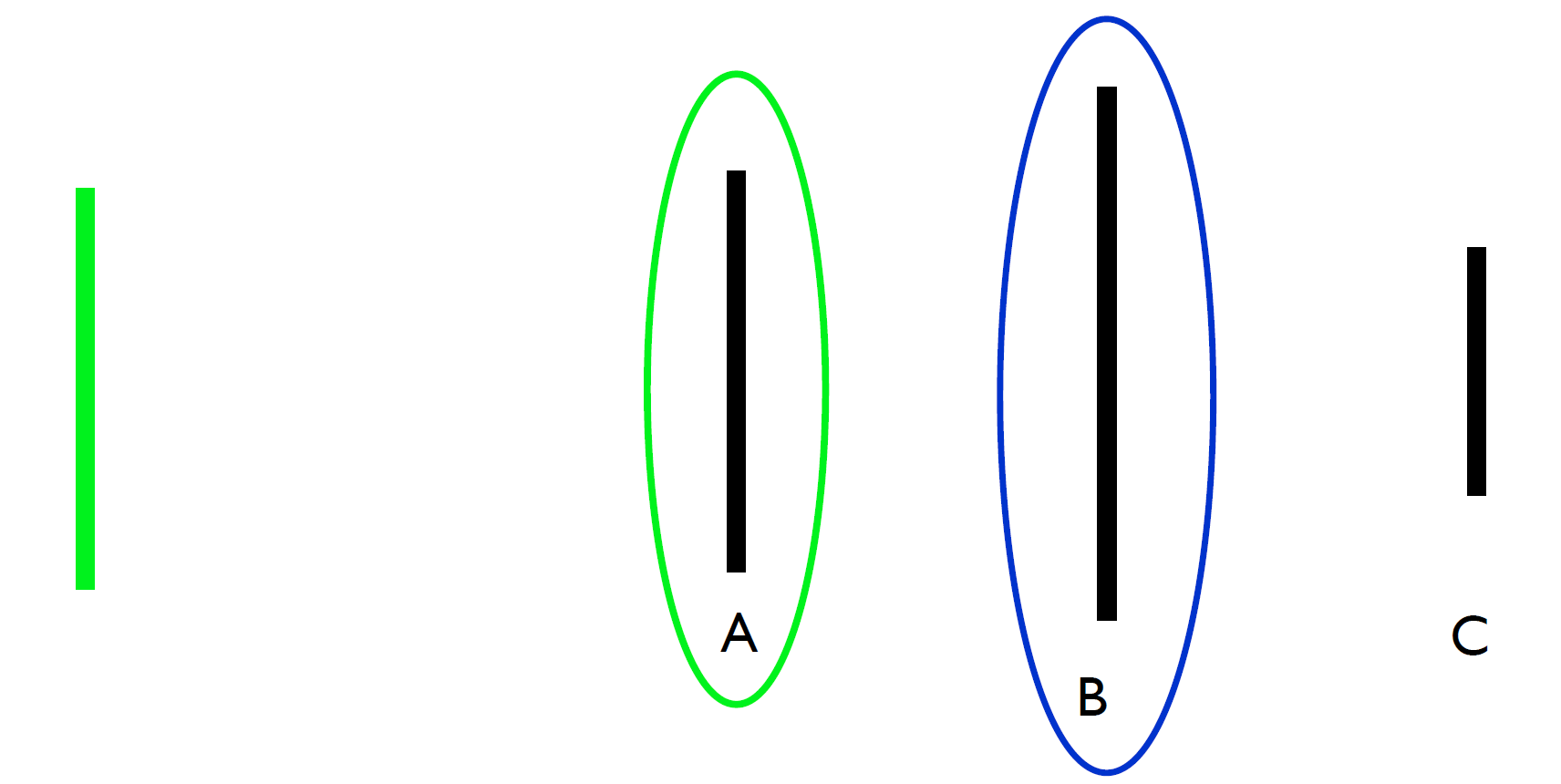}
	\caption{Example of the Asch experiment. The individuals openly discuss their individual beliefs as to which one of $A, B, C$ has the same length as the green line. Clearly $A$ is equal in length to the green line. The test individual is the red node.  The confederates (seven blue nodes) unanimously express belief in the same wrong answer, e.g. $B$.  }
	\label{fig:asch_group_example}
\end{figure}

In the experiment, eight individuals are instructed to judge a series of line lengths. Of the eight individuals, one is in fact the test subject, and the other seven ``confederates''\footnote{These other individuals have become referred to as ``confederates'' in later literature.} have been told a priori about what they should do. An example of the line length judging experiment is shown in Fig.~\ref{fig:asch_group_example}. There are three lines of unequal length, and the group has open discussions concerning which one of the lines $A,B,C$  is equal in length to the green line. Each individual is required to independently declare his choice, and the confederates (blue individuals) unanimously select the same wrong answer, e.g. $B$. The reactions of the test individual (red node) are then recorded, followed by a post-experiment interview to evaluate the test individual's private belief\footnote{In this section, we refer to $y_i, \hat{y}_i$ as beliefs, as the variables represent individual $i$'s certainty on an issue that is provably true or false. As noted in Section~\ref{sec:model}, our model is general enough to cover both subjective and intellective topics.}.
% Control groups without confederates were extremely successful (close to 100\% of groups) in determining the correct answer. 

In order to apply our model, and with Fig.~\ref{fig:asch_group_example} as an illustrative example, we frame $y_i, \hat{y}_i \in [0,1]$ to be individual $i$'s belief in the statement ``the green line is of the same length as line A." Specifically, $y_i = 1$ (respectively $y_i = 0$) implies individual $i$ is maximally certain the statement is true (respectively, maximally certain the statement is false). Asch found close to $100\%$ of individuals in control groups had $y_i(0) = 1$. Without loss of generality, we therefore denote the test individual as individual $1$ and set $y_1(0) = \hat{y}(0) = 1$. Confederates are set  to have $y_i(0) = \hat{y}_i(0) = 0$, for $i= 2, \hdots, n$, with $\lambda_i = 0$ and $\phi_i = 1$. That is, they consistently express maximal certainty that ``the green line of the same length as line A" is a \textbf{false} statement. 
%We initialise the test individual with $y_1(0) = \hat{y}_1(0) = 1$, i.e. the test individual initially has maximal certainty that ``the green line of the same length as line A" is \textbf{true}. 

It should be noted that in the experiments, Asch never assigned values of susceptibility $\lambda_i$, and resilience $\phi_i$ to the individuals because the quantitatively measured data by Asch was the number of incorrect answers over 12 iterations per group, and the behaviour of the individual being tested. However, based on his written description of individuals (including excerpts of the interviews), it was clear to the authors of this paper what the approximate range of values of the parameters $\lambda_i, \phi_i$ should be for each type of individual. (Some of these descriptions and excerpts will be provided immediately below). Also, the experiments did not attempt to determine the influence matrix $\mat{W}$  (at the time, influence network theory in the sense of DeGroot etc. had not yet been developed). The qualitative observations made in this section are invariant to the weights $w_{ij}$, and focus is instead placed on examining Asch's experimental results from the perspective of our model. In the following Section~\ref{ssec:SocialConcepts}, the impact of $\mat{W}$ (and in particular the weight $w_{11}$), and parameters $\phi_1$ and $\lambda_1$, are shown using analytic calculations.

\subsection{Types of Individuals}\label{sssec:SocialConcepts_types}

Asch observed three broad types of individuals. In particular, he divided the test individuals as: (i) independent individuals, (ii) yielding individuals with distortion of judgment, and (iii) yielding individuals with distortion of action. The assigned values for the parameters $\lambda_1$ and $\phi_1$ for each type of individual are summarised in Table.~\ref{tab:asch_table}. Values of $\phi_1, \lambda_1$ in this neighbourhood generate responses that are qualitatively the same at a high level; the differences lie in the exact values of the final opinions.

\emph{Independent} individuals can be divided further into different subgroups depending on the reasoning behind their independence, but this will not be considered because we focus only on the final outcome or observed result and not the reasons for independence. Asch identified an independent individual as someone who was strongly confident that $A$ was correct. This individual did not change his expressed belief, i.e. did not yield to the confederates' unanimous declaration that $A$ was incorrect, despite the confederates insistently questioning the individual. Asch's descriptions indicate that the test individual is extremely stubborn (i.e. closed to influence) and confident his belief is correct, and is resilient to the group pressure. It is then obvious that one would assign to such individuals values of $\lambda_1$ close to zero and $\phi_1$ close to one. With the framing of the experiments given above, our model would be said to accurately capture an independent individual if test individual $1$ with parameter values of $\lambda_1$ close to zero and $\phi_1$ close to one, has final beliefs $\hat{y}_1^*, y_1^* \approx 1$.

%Some showed \emph{distortion of judgment/perception}, where individual $1$ expressed and held a private belief in line with the confederate majority, . Others showed \emph{distortion of action} whereby the test individual expressed a belief similar to the confederate majority, but in the post-experiment interview maintained private belief of the correct answer; in our model we would . Based on Asch's qualitative descriptions of these 3 types of individuals \cite{asch1951group_pressure_effects}, 

Asch also identified \emph{yielding} individuals, who could be divided into two groups. Those who experienced a distortion of \emph{judgment/perception} either (i) lacked confidence, assumed the group was correct and thus concluded $A$ was incorrect, or (ii) did not realise he had been influenced by the group at all and changed his private belief to be certain that $A$ was incorrect. This indicates that the individual is open to influence (i.e. not stubborn in $y_1(0) = 1$) and is highly affected by the group pressure (i.e. not resilient). One concludes that for such individuals $\lambda_1$ is likely to be close to one, and $\phi_1$ to be close to zero. As shown in the sequel, it turns out that the value of $\phi_1$ plays only a minor role for such an individual because he is already extremely susceptible to influence. For our model to accurately capture such an individual, then for $\lambda_1$ close to one, and $\phi_1$ close to zero, one expects $y_1^*, \hat{y}_1^* \approx 0$. 

Other yielding individuals experienced a \emph{distortion of action}. This type of individual, on being interviewed (and before being informed of the true nature of the experiment) stated that he remained privately certain that $A$ was the correct answer, but suppressed his observations as to not publicly generate friction with the group. Such an individual has full \emph{awareness of the difference} between the truth and the majority's position. This individual is closed to influence (i.e. stubborn) but not resilient, and it is predicted that such individuals will have $\lambda_1$ and $\phi_1$ both close to zero. If our model were to accurately capture such an individual, then the final beliefs would be expected to be $y_1^* \approx 1$ and $\hat{y}_1^* \approx 0$.

\begin{table}
	\centering
	\caption{Types of test individuals and their susceptibility and resilience parameters}\label{tab:asch_table}
	\begin{tabular}{ l l l}
		& $\lambda_1$ & $\phi_1$ \\ 
		\hline 
		\textbf{Independent} &  low & high \\ 
		\hline 
		\textbf{Yielding, judgment distortion} & high  & any  \\ 
		\hline 
		\textbf{Yielding, action distortion} & low & low \\ 
		\hline 
	\end{tabular} 
\end{table}

\subsection{Theoretical Analysis}\label{ssec:SocialConcepts}

This section will present theoretical calculations of Asch's experiments in the framework of the our model, showing how $y_1, \hat{y}_1$ vary with $\mat{W}$, $\lambda_1 \in [0,1]$ and $\phi_1 \in [0,1]$. Analysis will be conducted for $n \geq 2$, to investigate the effects of the majority size on the belief evolution. We make the mild assumption that $w_{11} \in (0,1)$, which implies that individual $1$ considers his/her own private belief during the discussions.

Because $\lambda_i = 0$ and $\phi_i =1$ for all $i = 2, \hdots, n$, one concludes from \eqref{eq:private_op} and \eqref{eq:public_op} that $y_i(t) = \hat{y}_i(t) = 0$ for all $t$. With $\vect{y}(0) = [1, 0, \hdots, 0]^\top$, it follows that test individual $1$'s belief evolves as
\begin{align}\label{eq:asch_test_indiv_system}
\begin{bmatrix} y_1(t+1) \\ \hat{y}_1(t) \end{bmatrix} & = \mat{V}\begin{bmatrix} y_1(t) \\ \hat{y}_1(t-1) \end{bmatrix} + \begin{bmatrix} 1-\lambda_1 \\ 0 \end{bmatrix}.
\end{align}
where
\begin{equation}
\mat{V} = \begin{bmatrix} \lambda_1 w_{11} & 0 \\ \phi_1 & \frac{1}{n}(1-\phi_1) \end{bmatrix}.
\end{equation}
From the fact that $n \geq 2$, $\lambda_1 \in [0,1]$, $w_{11} \in (0,1)$, and $\phi_1 \in [0,1]$, it follows that $\mat{V}$ has eigenvalues inside the unit circle and thus the system in \eqref{eq:asch_test_indiv_system} converges to limit exponentially fast. Straightforward calculations show that this limit is given by 
\begin{align}
\lim_{t\to\infty} y_1(t) \triangleq y_1^* & = \frac{1-\lambda_1}{1-\lambda_1 w_{11}} \\
\lim_{t\to\infty} \hat{y}_1(t) \triangleq \hat{y}_1^* & = \frac{n\phi_1}{n-1 + \phi_1} y_1^*.
\end{align}

\begin{figure}
	\centering
	\begin{minipage}{0.8\linewidth}
		\includegraphics[height=\linewidth,angle=-90]{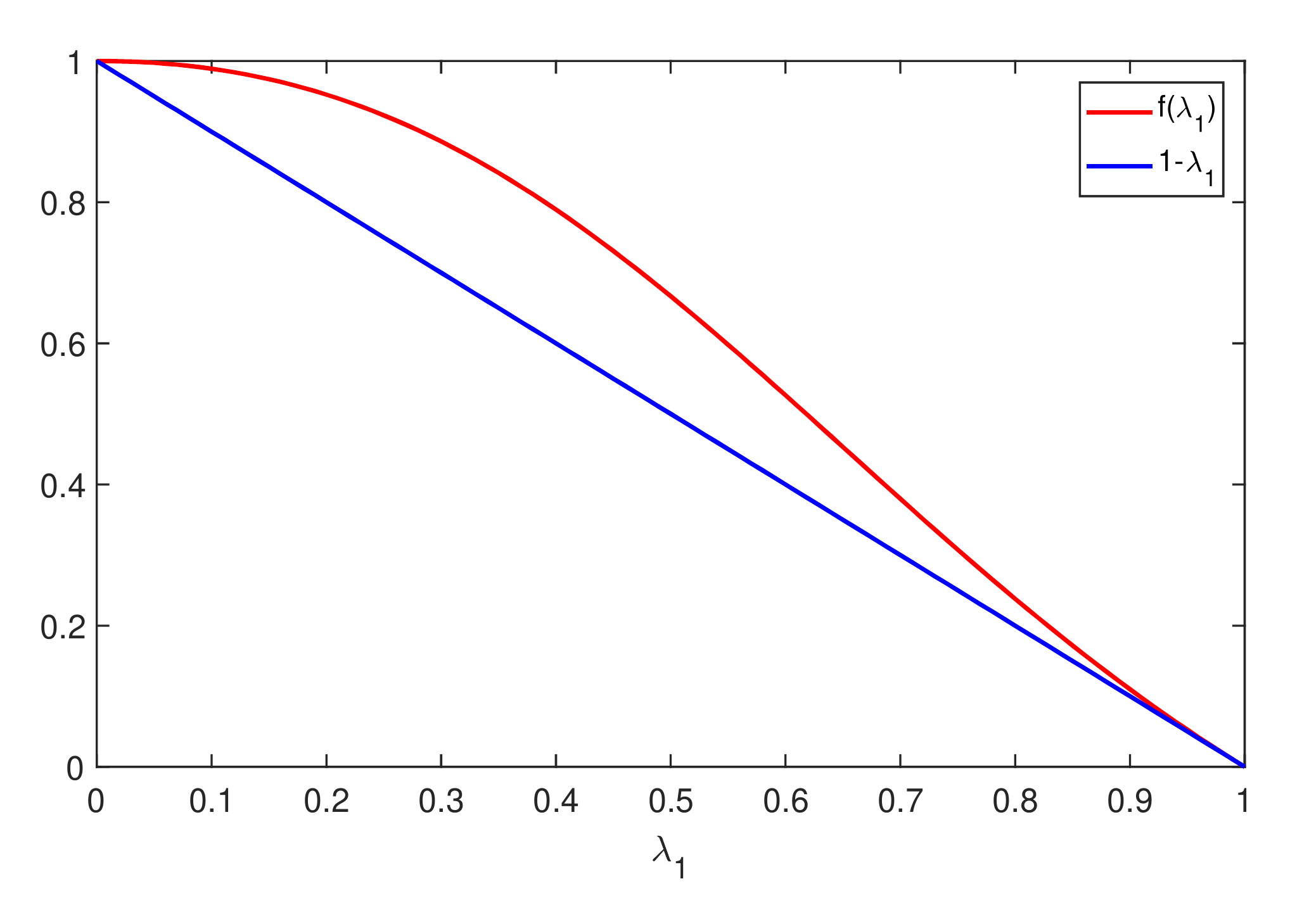}
		\caption{The function $f(\lambda_1)$ and $1-\lambda_1$ plotted against $\lambda_1$. The analytical calculations show that $y_1^* = f(\lambda_1)$, and thus the red line represents individual $1$'s final private belief as a function of his susceptibility to influence.}
		\label{fig:f_function}
	\end{minipage}
	\vfill
	\begin{minipage}{0.8\linewidth}
		\includegraphics[height=\linewidth,angle=-90]{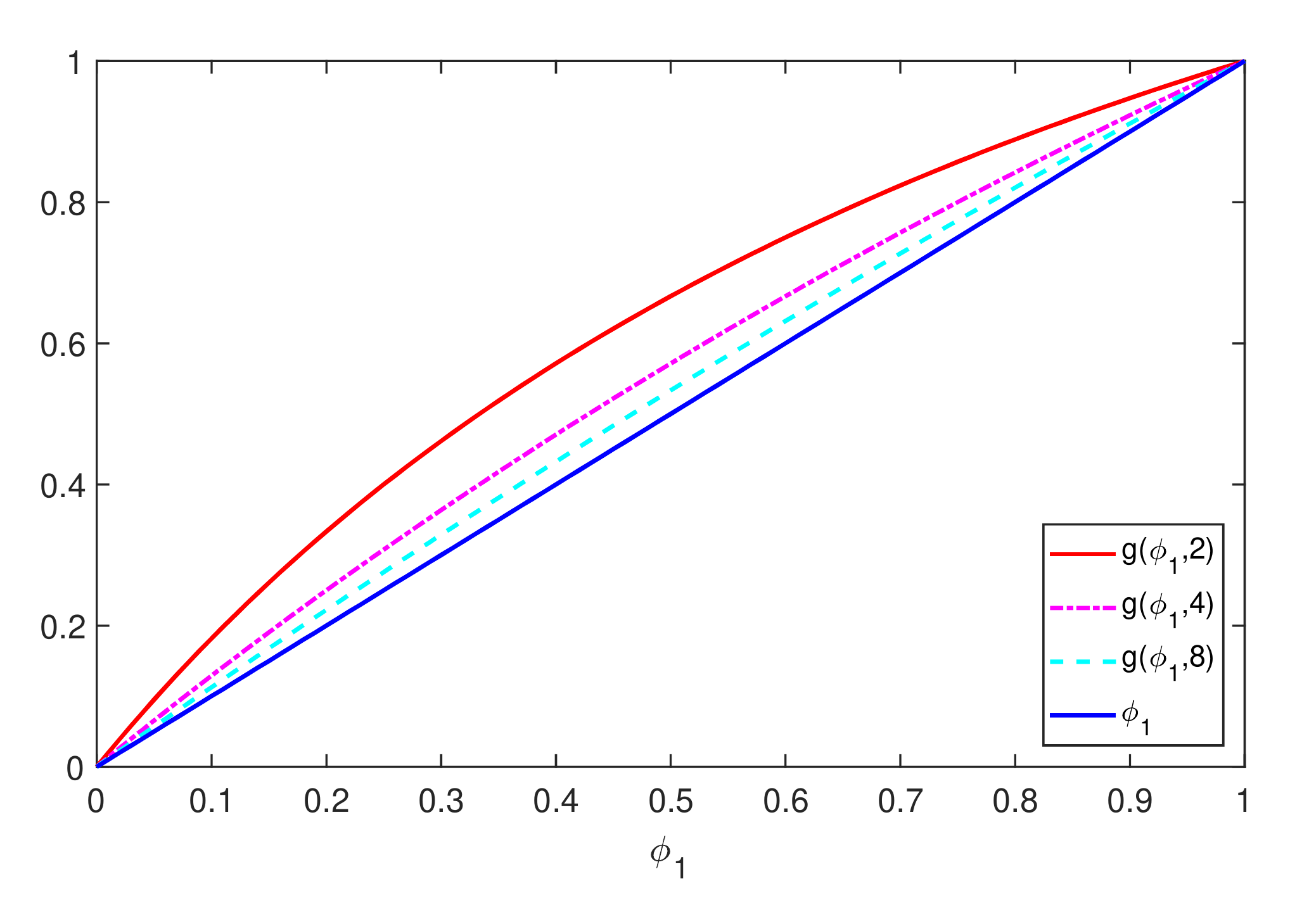}
		\caption{The function $g(\phi_1,n)$, with $n=2,4,8$, plotted against $\phi_1$. The analytical calculations show that $\hat{y}_1^* = g(\phi_1,n)y_1^*$, and thus the plot shows how the test individual's final expressed opinion is changed from his final private opinion by his resilience $\phi_1$, and by $n$. }
		\label{fig:g_function}
	\end{minipage}
\end{figure}

From this, one concludes that the test subject's final private belief is dependent on his level of stubbornness in believing that $A$ is the correct answer, i.e. $\lambda_1$, and on his self-weight $w_{11}$, i.e. how much he trusts his own belief relative to the others in the group. Interestingly, $y_1^*$ does not depend on individual $1$'s resilience $\phi_1$, though it must be noted that this is a special case when the other individuals are all confederates. In general networks beyond the Asch framework, $y_1^*$ will depend not only on $\phi_1$, but also the other $\phi_i$. For simplicity, consider a natural selection of $w_{ii} = 1 - \lambda_i$ \cite{friedkin1990_FJsocialmodel}.  As a result, one obtains that $y_1^* = (1-\lambda_1)/(1-\lambda_1 (1-\lambda_1))$. Examination of the function $f(\lambda_1) = (1-\lambda_1)/(1-\lambda_1 (1-\lambda_1))$, for $\lambda_1 \in [0,1]$, reveals how the test subject's final private belief changes as a function of his openness to influence; the function $f(\lambda_1)$ is plotted in Fig.~\ref{fig:f_function}. Notice that $f(\lambda_1) = (1-\lambda_1)/(1-\lambda_1 (1-\lambda_1)) \geq 1 - \lambda_1$ for $\lambda_1 \in [0,1]$ with equality if and only if $\lambda_1 = \{0,1\}$. This implies that the test individual's final $y_1^*$ will always be greater than his stubbornness $1-\lambda_1$, except if he has $\lambda_1 = 0$ (maximally stubborn) or $\lambda_1 = 1$ (maximally open to influence). 

Next, consider the final expressed belief, which is given as $\hat{y}_1^* = \frac{n\phi_1 }{n-1 + \phi_1}y_1^*$. The relative closeness of $\hat{y}_1^*$ to $y_1^*$, as measured by $\hat{y}_1^*/y_1^*$, is determined by $n$ and $\phi_1$. Define $g(\phi_1,n) = \frac{n\phi_1}{n-1 + \phi_1}$. The function $g(\phi_1,n)$ is plotted in Fig.~\ref{fig:g_function}. Observe that $g(\phi_1,n) \geq  \phi_1$ for any $n$, for all $\phi_1 \in [0,1]$, and with equality if and only if $\phi_1 = \{0,1\}$. This implies that the test individual's final expressed belief will always be closer to his final private belief than his resilience level. Most interestingly, observe that $g(\phi_1,n) \to \phi_1$ from above, as $n \to \infty$, but the difference between $g(\phi_1,n)$ and $\phi_1$ when going from $n = 2$ to $n = 2\times 2 = 4$ is much greater than the differences going from $n = 4$ to $n = 4\times 2 = 8$.  This may explain the observation in \cite{asch1951group_pressure_effects} that increasing the majority size did not produce a correspondingly larger distortion effect beyond majorities of three to four individuals, at least for test individuals with low $\lambda_1$. That is, an increase in $n$ does not produce a matching increase in distortion of the final expressed opinion from the final private opinion, represented as $\hat{y}_1^*/y_1^* = g(\phi_1,n) \to 1$ as $n\to \infty$.

Also of note is that for individuals with $\lambda_1$ close to one, $y_1^*$ is already close to zero, and bounds $\hat{y}_1^*$ from above. The \emph{magnitude of the difference}, $\vert y_1^* - \hat{y}_1^*\vert$, only changes slightly as $\phi_1$ is varied, which indicates that for individuals who yielded with distortion of judgment, the value of $\phi_1$ plays only a minor role in the determining the absolute (as opposed to relative) difference between expressed and private beliefs. This is in contrast to individuals with low susceptibility, where the behaviour of an individual can vary significantly by varying $\phi_1$ from 1 to 0. 

\subsection{Simulations}\label{ssec:asch_sim}

\begin{figure*}
	\begin{minipage}{0.325\linewidth}
	\begin{subfigure}[t]{\textwidth}
		\includegraphics[height=\linewidth,angle=-90]{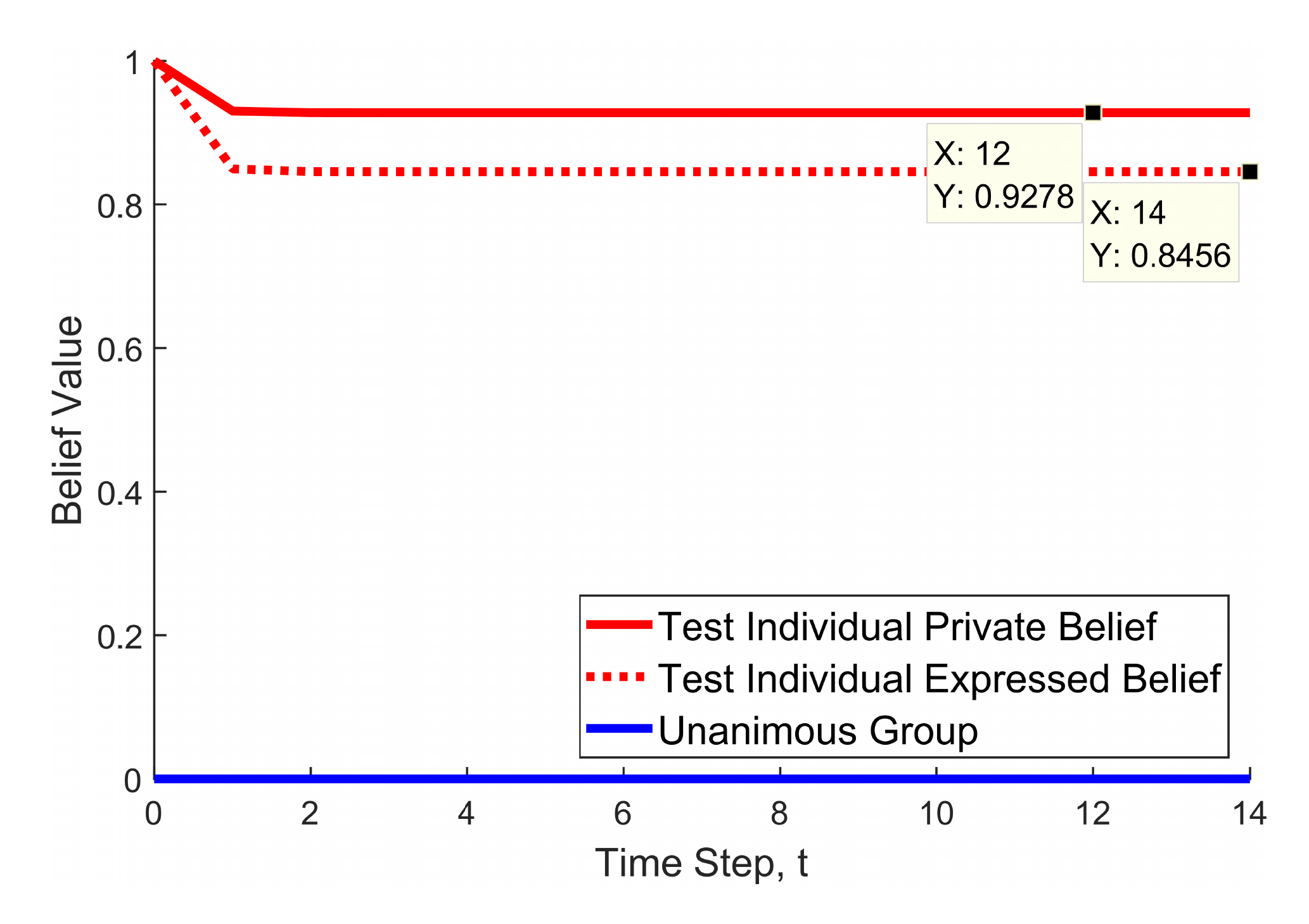}
		\caption{An \emph{independent} individual,  with $\lambda_1 = 0.1, \phi_1 = 0.9$.  }
		\label{fig:independent_individual}
	\end{subfigure}
\end{minipage}
\hfill
	\begin{minipage}{0.325\linewidth}
	\begin{subfigure}[t]{\textwidth}
		\includegraphics[height=\linewidth,angle=-90]{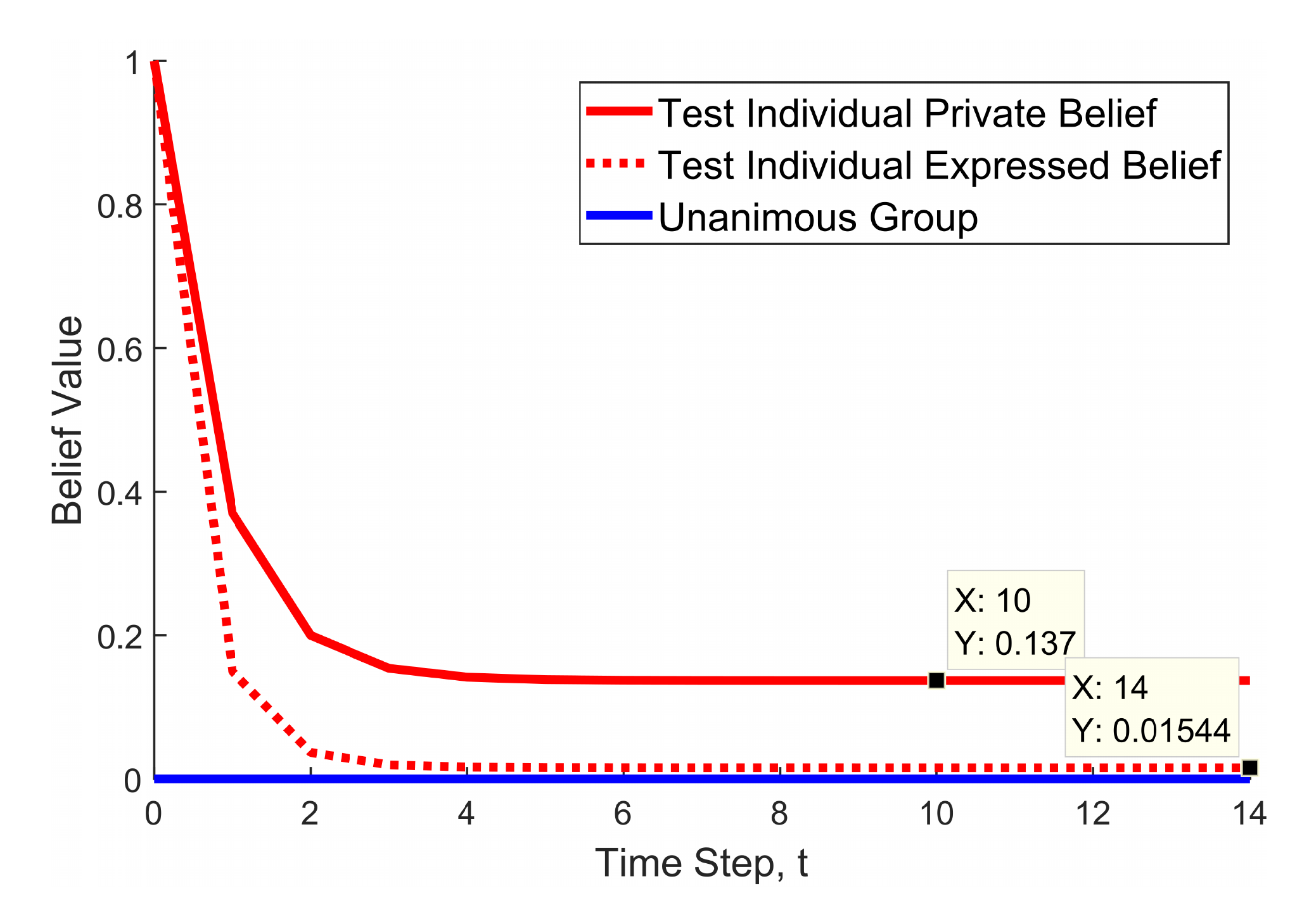}
		\caption{A \emph{yielding} individual with \emph{distortion of judgment},  with $\lambda_1 = 0.9, \phi_1 = 0.1$.  }
		\label{fig:distortion_judgement}
	\end{subfigure}
\end{minipage}
	\hfill
\begin{minipage}{0.325\linewidth}
	\begin{subfigure}[t]{\textwidth}
		\includegraphics[height=\linewidth,angle=-90]{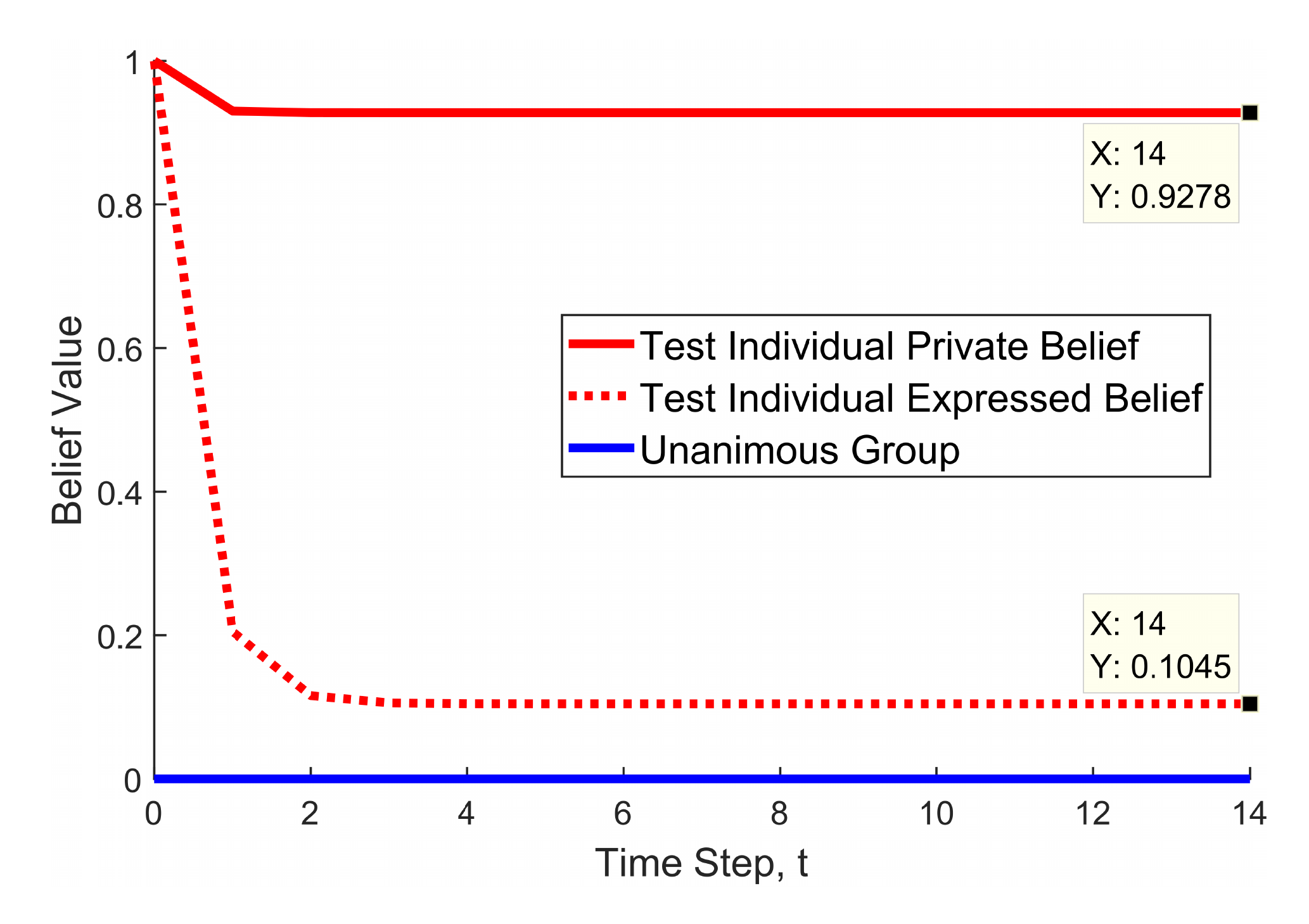}
		\caption{A \emph{yielding} individual with \emph{distortion of action}, with $\lambda_1 = 0.1, \phi_1 = 0.1$.  }
		\label{fig:distortion_action}
	\end{subfigure}
\end{minipage}
	\caption{Fig.~ \ref{fig:independent_individual}, \ref{fig:distortion_action}, and \ref{fig:distortion_judgement} show the evolution of beliefs for all three types of reactions recorded by Asch, as they appear in our model. The red solid and dotted line denote the private and expressed belief, respectively, of the test individual $1$ (i.e. $y_1(t)$ and $\hat{y}_1(t)$). The blue line is the belief of the unanimous confederate group, who express a belief of $\hat{y}_i(t) = 0$.}
\label{fig:asch_example}
\end{figure*}

The Asch experiments are simulated using the proposed model. An arbitrary $\mat{W}$ is generated with weights $w_{ij}$ sampled randomly from a uniform distribution and normalised to ensure $\sum_{j=1}^n w_{ij} = 1$. The other parameters are described in the third paragraph of Section~\ref{sec:asch}.
%As above, individual $1$ is set as the test subject, with individuals $2, \hdots, 8$ being the 7 confederates in an $n=8$ network. The initial conditions are $\vect{y}(0) = \hat{\vect y}(0) = [1, 0, \hdots, 0]^\top$, while $\lambda_i = 0$ and $\phi_i = 1$ for $i = 2, \hdots, 8$. 
In the following plots of Fig.~\ref{fig:independent_individual}, \ref{fig:distortion_judgement} and \ref{fig:distortion_action}, the values of $\lambda_1$ and $\phi_1$ are given. The red lines correspond to test individual $1$, with the solid line showing private belief $y_1(t)$ and the dotted line showing expressed belief $\hat{y}_1(t)$. The blue line represents the confederates $k=2, \hdots, 8$, who have $y_k(t) = \hat{y}_k(t) = 0$ for all $t$. 

Figure~\ref{fig:independent_individual} shows the evolution of beliefs when the test individual is \emph{independent}. It can be seen that both the private and expressed beliefs of $v_1$ are largely unaffected by the confederates' unanimous expressed belief and the pressure exerted by the group. Note that $\hat{y}_1^* < y_1^*$, which is also reported in \cite{asch1951group_pressure_effects}; despite expressing his belief that $A$ is the correct answer, one independent test individual stated ``You're \emph{probably} right, but you may be wrong!", which might be seen as a concession towards the majority belief. There is also a small shift away from maximal certainty of $y_i = 1$, with $y_1^* \approx 0.93$; in \cite{asch1951group_pressure_effects}, one independent test individual stated 
\begin{quote}\emph{I would follow my own view, though part of my reason would tell me that I might be wrong.}\end{quote}

Figure~\ref{fig:distortion_judgement} shows the belief evolution of a yielding test individual who, under group pressure, exhibits \emph{distortion of judgment/perception}. The figure shows that both $y_1^*$ and $\hat{y}_1^*$ are heavily influenced by the group pressure, and thus individual $1$ is no longer privately certain that $A$ is the correct answer. In other words, this individual is highly susceptible to interpersonal influence, and even his private view becomes affected by the majority. Of great interest is the evolution of beliefs observed in Fig.~\ref{fig:distortion_action}, which involves an experiment with a yielding test individual exhibiting \emph{distortion of action}. According to Asch, Individual $1$
\begin{quote}
	\emph{yields because of an overmastering need to not appear different or inferior to others, because of an inability to tolerate the appearance of defectiveness in the eyes of the group} \textasciitilde \cite{asch1951group_pressure_effects}.
\end{quote}
In other words $v_1$'s expressed belief $y_1^*$ is heavily distorted by the pressure to conform to the majority. However, this individual is still able to ``conclude that they [themselves] are not wrong'' \cite{asch1951group_pressure_effects}, i.e. $y_i^* \approx 0.93$.

Other simulations with values of $\lambda_1,\phi_1$ in the neighbourhood of those used also display similar behaviour as shown in Fig.~\ref{fig:independent_individual} to \ref{fig:distortion_judgement}, indicating a robust ability for our model to capture Asch's experiments is an intrinsic property of the model, and rather than resulting from careful reverse engineering. All three types of individual behaviours can be predicted by our model using pairs of parameters $\lambda_i, \phi_i$, providing a measure of validation for our model. At the same time, we have provided an agent-based model explanation of the empirical findings of Asch's experiments; it might now be possible to analyse the many subsequent works derived from Asch can be analysed common framework, whereas existing static models of conformity are tied to specific empirical data (see the Introduction). The Friedkin--Johnsen model has also been applied to the Asch experiments  \cite{friedkin2011social_book}, but (unsurprisingly) was not able to capture all of the types of individuals reported because the Friedkin--Johnsen model does not assume that each individual has a separate private and expressed belief. 

\subsection{Threshold Variant and Asch's Second Experiments}\label{ssec:threshold}
The simulations above assumed that the individuals express a continuous real-valued beliefs $\hat y_i(t)$, whereas it is perhaps more appropriate to set $\hat y_i(t)$ as a binary variable, with $\hat y_i(t) = 1$ and $\hat y_i(t) = 0$ denoting individual $i$ picking $A$ and not picking $A$ as the correct answer. The proposed model can be modified to accommodate situations where the expressed variable denotes an action, or decision by replacing \eqref{eq:public_op} with
\begin{equation}\label{eq:thres_public_op}
\hat y_i(t) = \sigma_i\left(\phi_i y_i(t) + (1-\phi_i)\hat{y}_{i,\text{lavg}}(t-1)\right),
\end{equation}
where $\sigma_i(x) : [0,1] \to \{0, 1\}$ is a threshold function satisfying $\sigma_i(x) = 0$ if $x \in[0, \tau_i]$ and $\sigma_i(x) = 1$ if $x\in (\tau_i, 1]$, for some threshold value $\tau_i \in (0,1)$. Applying the threshold variant of the model with $\tau_i = 0.5$ yields no qualitative difference for the simulations in Section~\ref{ssec:asch_sim}. That is, pairs of parameter values $\lambda_1, \phi_1$ which in the original model were associated with an independent, distortion of action, or distortion of judgment individual (Table~\ref{tab:asch_table}) were almost always also associated with the same type of individual in the threshold model.

\subsubsection{Calculations}\label{sssec:calc_thres}

Because of the highly specialised setup for the Asch experiments, it turns out that one can theoretically calculate the final beliefs of test individual 1 even under the threshold model. This would not be the case for the threshold model in general scenarios. In fact, it is unclear if the threshold model will always converge in a general setting, especially if individuals update synchronously. 

We perform calculations for Asch's experiments (Section~\ref{sec:asch}). First, we remark that the private opinion dynamics $y_1(t)$ of test individual 1 is unchanged in the threshold model when compared to the original model, since the expressed beliefs of all of individual 1's neighbours are stationary.  Thus, $\lim_{t\to\infty} y_1(t) \triangleq y_1^* = \frac{1-\lambda_1}{1-\lambda_1 w_{11}}$ as in the original model calculations in Section~\ref{ssec:SocialConcepts}.  

One can then consider $y_1(t)$ as an input to \eqref{eq:thres_public_op}. It follows that $\hat y_1(t)$ converges. In particular, and assuming global public opinion is used, then $\lim_{t\to\infty} \hat y_1(t) \triangleq \hat y_1^* = 1$ if $\phi_1 y_1^* + (1-\phi_1)\frac{1}{n} \geq \tau_1$ and $ \hat y_1^* = 0$ if $\phi_1 y_1^* < \tau_1$. There is a small interval region $\tau_1 \in \left(\phi_1 y_1^*, \phi_1 y_1^* + (1-\phi_1)\frac{1}{n}\right)$ of width $(1-\phi_1)/n$ where $\hat y_1^*$ depends on the initial condition $\hat y_1(0)$.

\subsubsection{Asch's Second Experiments}\label{sssec:asch_2nd}

Asch conducted several variations to the original experiments, as reported in \cite{asch1951group_pressure_effects,asch1955opinions}. In one particular variation, one confederate also told a priori to select the correct answer; the frequency of individuals showing distortion of action or distortion of judgment decreased dramatically. We now frame this variation of the experiment in our model's framework, and call it Asch's Second Experiment for convenience. The parameter matrix $\mat W$, and parameters $\lambda_i$ and $\phi_i$, $i = 1, \hdots, 8$ are unchanged from the first experiment described in Section~\ref{sec:asch}. The setup of individual $1$ is also the same.  However, different from Section~\ref{sec:asch}, the $n-1$ confederates' beliefs are now set to be $y_2(0) = \hat y_2(0) =1$, and $y_i(0) = \hat y_i(0) = 0$ for $i = 3, \hdots, n$. It should be noted that theoretical calculations of the final private and expressed beliefs of individual $1$ can also be completed, following the same method as in Section~\ref{sssec:calc_thres}.

\subsubsection{Simulations}
We now provide simulations for Asch's Second Experiment, using both the original model proposed in \eqref{eq:public_op}, and the threshold model in \eqref{eq:thres_public_op}.

\textit{Case 1:} The behaviour of individuals with high $\phi_1$  and low $\lambda_i$ (independent individuals in Asch's First Experiment) are the same, qualitatively, when comparing the original model and the threshold model. We omit the simulation results for such individuals.

\begin{figure*}
	\begin{minipage}{0.375\linewidth}
		\centering
		\begin{subfigure}[t]{\textwidth}
			\includegraphics[height=\linewidth,angle=-90]{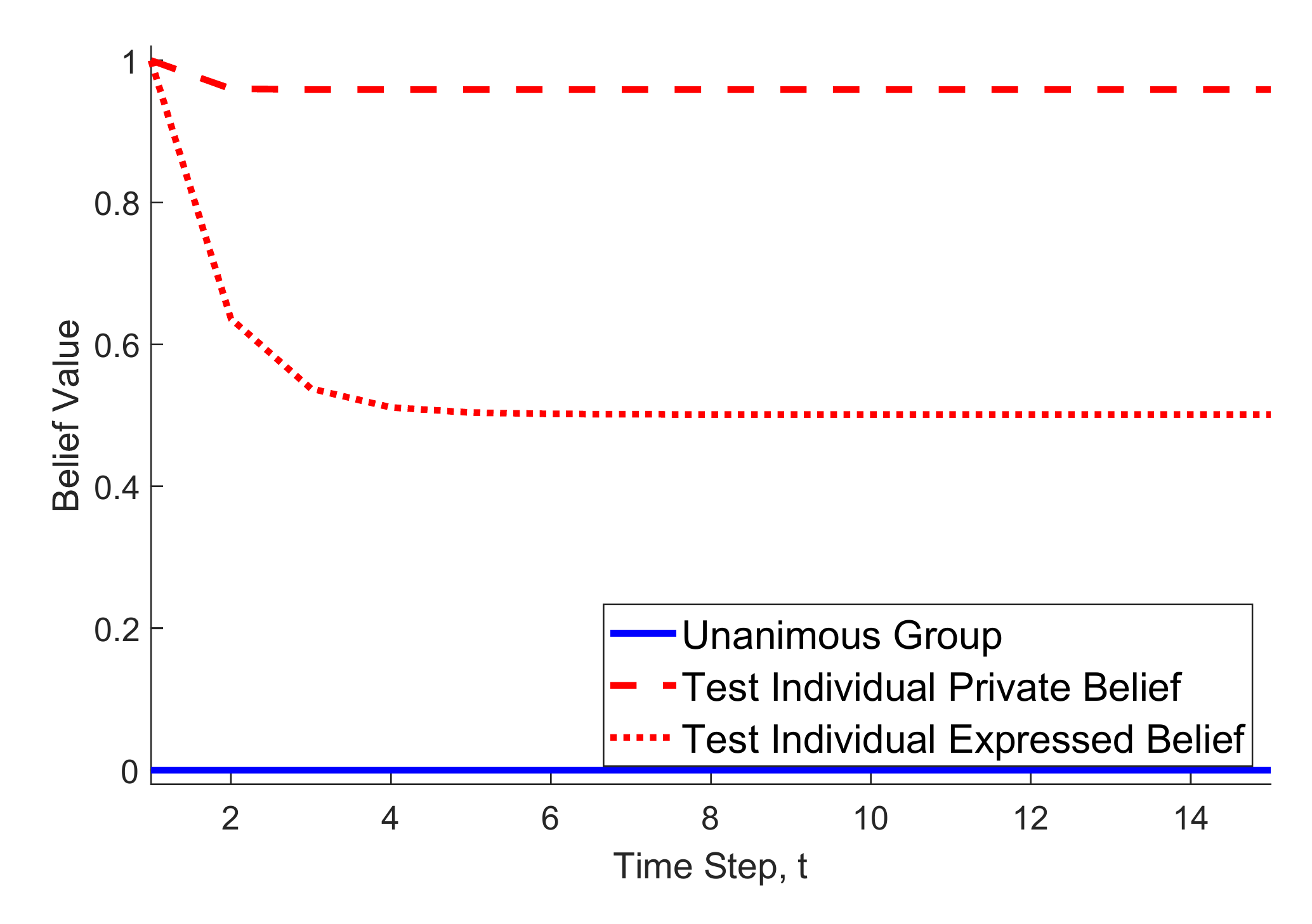}
			\caption{Asch's second experiment: An individual with $\lambda_1 = 0.1, \phi_1 = 0.1$, original model.  }
			\label{fig:distortion_action_v2_OG}
		\end{subfigure}
	\end{minipage}
	\hfill
	\begin{minipage}{0.375\linewidth}
		\centering
		\begin{subfigure}[t]{\textwidth}
			\includegraphics[height=\linewidth,angle=-90]{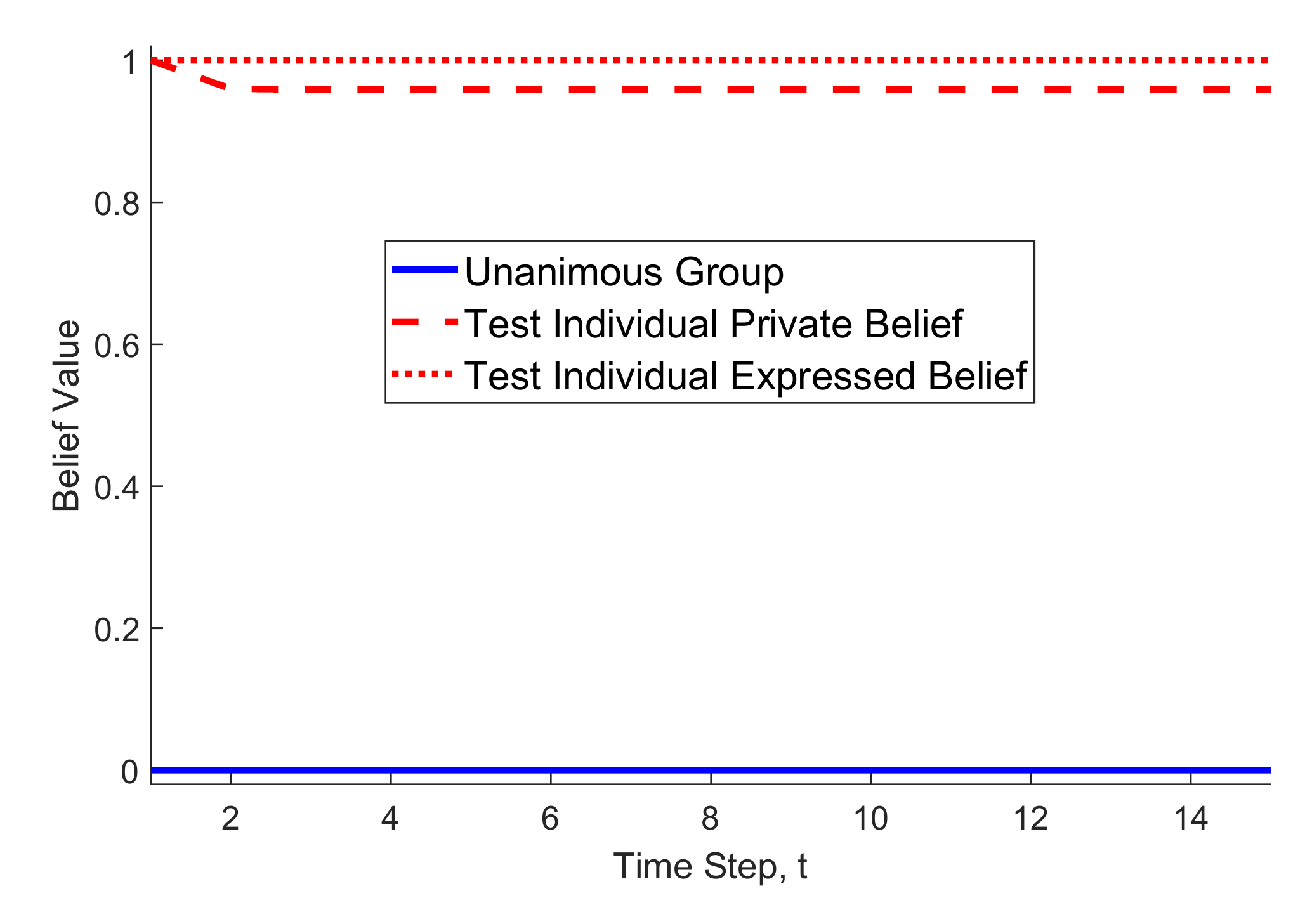}
			\caption{Asch's second experiment: An individual with $\lambda_1 = 0.1, \phi_1 = 0.1$, threshold model.  }
			\label{fig:distortion_action_v2_TH}
		\end{subfigure}
	\end{minipage}
	\caption{Fig.~\ref{fig:distortion_action_v2_OG} and \ref{fig:distortion_action_v2_TH} show the evolution of beliefs, for two different models, of the variation of the Asch experiment where a second actor supports the truth. The red dashed and dotted line denote the private and expressed belief, respectively, of the test individual $1$ (i.e. $y_1(t)$ and $\hat{y}_1(t)$). The blue line is the belief of the unanimous confederate group, who express a belief of $\hat{y}_i(t) = 0$.}
\end{figure*}

\textit{Case 2:} Next, we simulate a test individual that has low $\phi_1$  and low $\lambda_i$ (in Asch's First Experiment, these individuals were said to show distortion of action). Fig.~\ref{fig:distortion_action_v2_OG} and \ref{fig:distortion_action_v2_TH} show a test individual with $\lambda_1 = 0.1, \phi_1 = 0.1$, for the original and threshold model, respectively. 
%It can be seen that in the original model, the expressed belief $y_1^*$ of test individual $1$ is just above $0.5$, which shows that the truth-telling confederate does indeed have an effect. The effect is much more significant

\begin{figure*}
	\begin{minipage}{0.375\linewidth}
		\centering
		\begin{subfigure}[t]{\textwidth}
			\includegraphics[height=\linewidth,angle=-90]{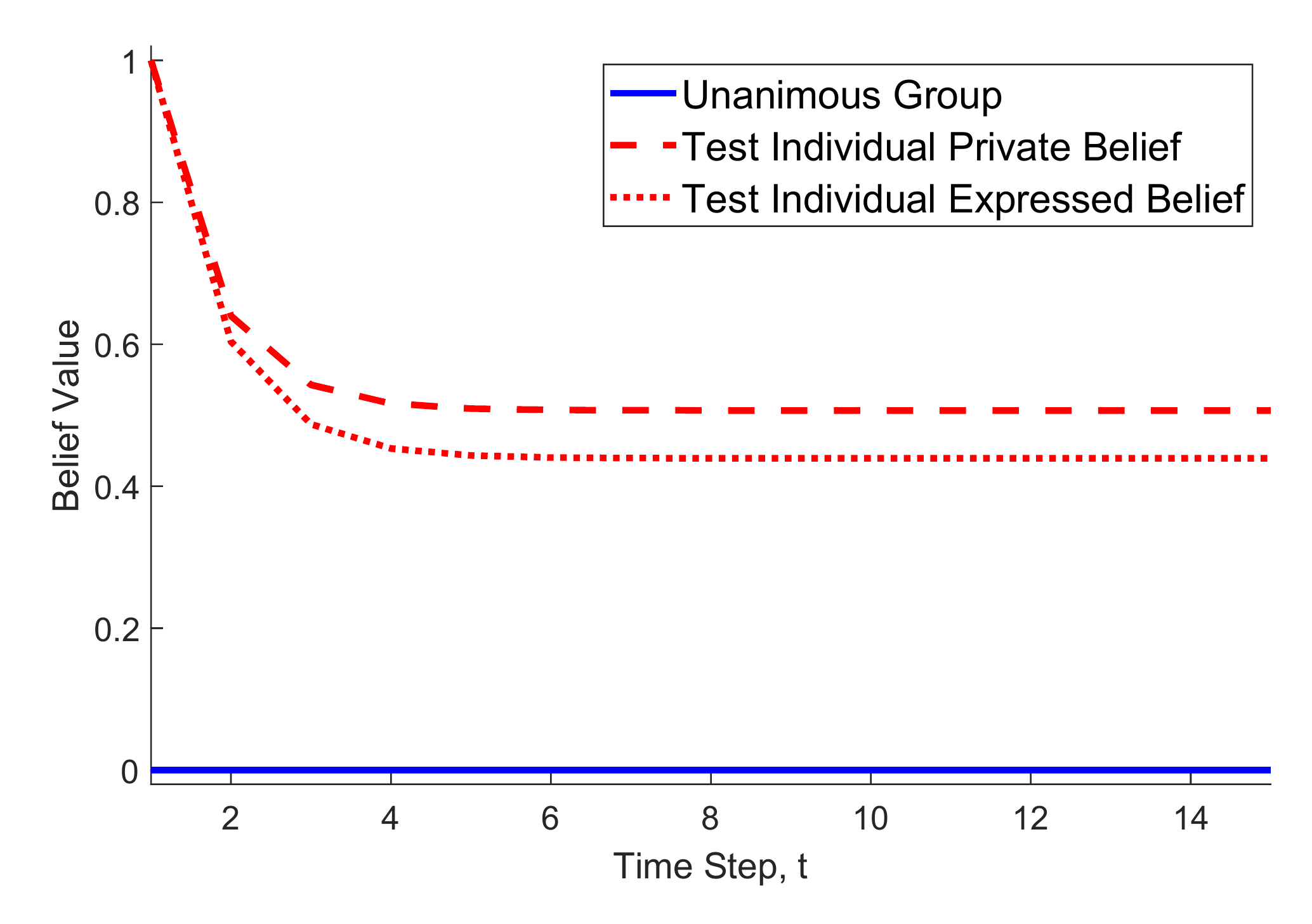}
			\caption{Asch's second experiment: An individual with $\lambda_1 = 0.9, \phi_1 = 0.1$, original model.  }
			\label{fig:distortion_judgment_v2_OG}
		\end{subfigure}
	\end{minipage}
	\hfill
	\begin{minipage}{0.375\linewidth}
		\centering
		\begin{subfigure}[t]{\textwidth}
			\includegraphics[height=\linewidth,angle=-90]{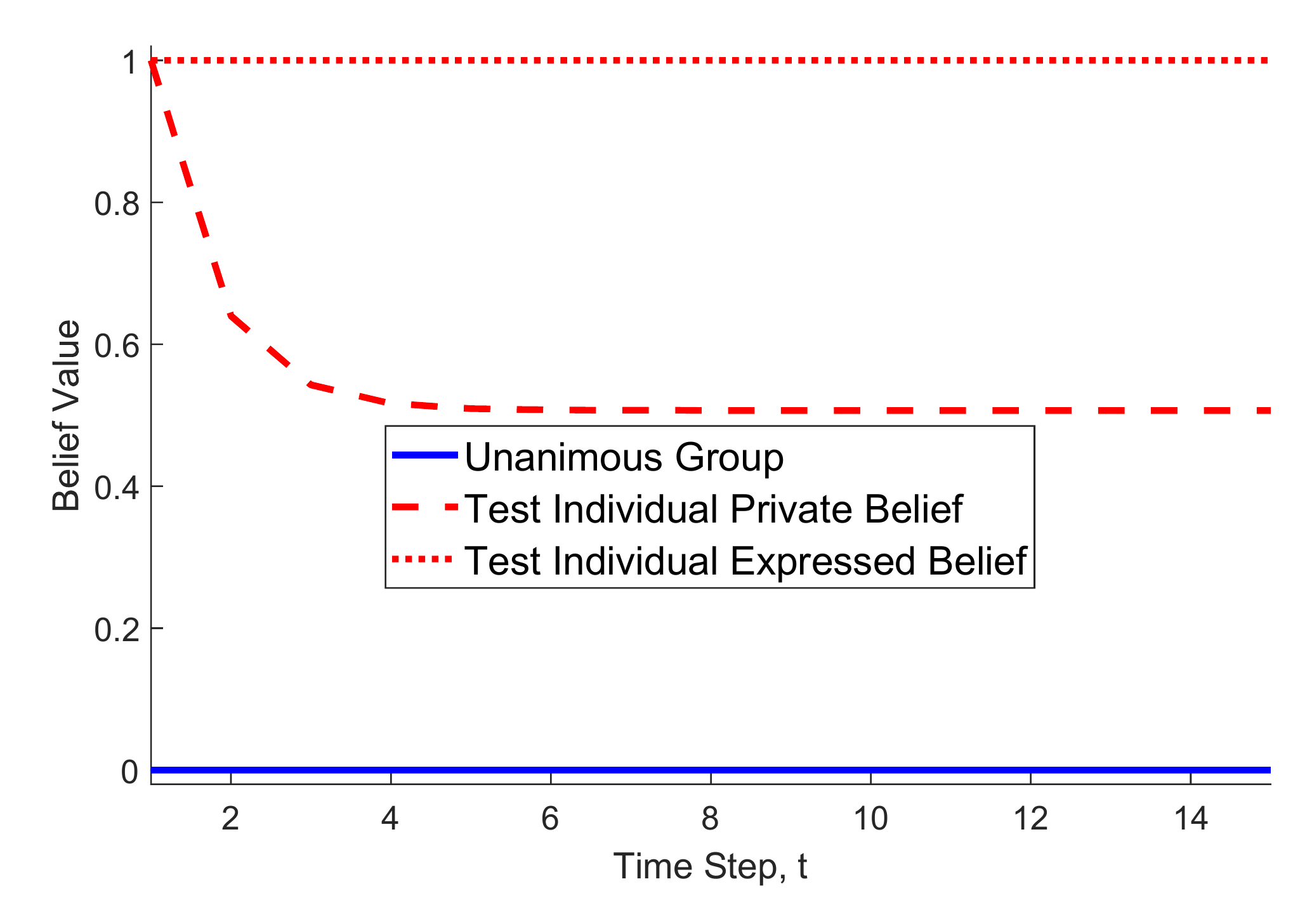}
			\caption{Asch's second experiment: An individual with $\lambda_1 = 0.9, \phi_1 = 0.1$, threshold model.  }
			\label{fig:distortion_judgment_v2_TH}
		\end{subfigure}
	\end{minipage}
	\caption{Fig.~\ref{fig:distortion_judgment_v2_OG} and \ref{fig:distortion_judgment_v2_TH} show the evolution of beliefs, for two different models, of the variation of the Asch experiment where a second actor supports the truth. The red dashed and dotted line denote the private and expressed belief, respectively, of the test individual $1$ (i.e. $y_1(t)$ and $\hat{y}_1(t)$). The blue line is the belief of the unanimous confederate group, who express a belief of $\hat{y}_i(t) = 0$.}
\end{figure*}

\begin{figure}
	\centering
	\includegraphics[height=0.8\linewidth,angle=-90]{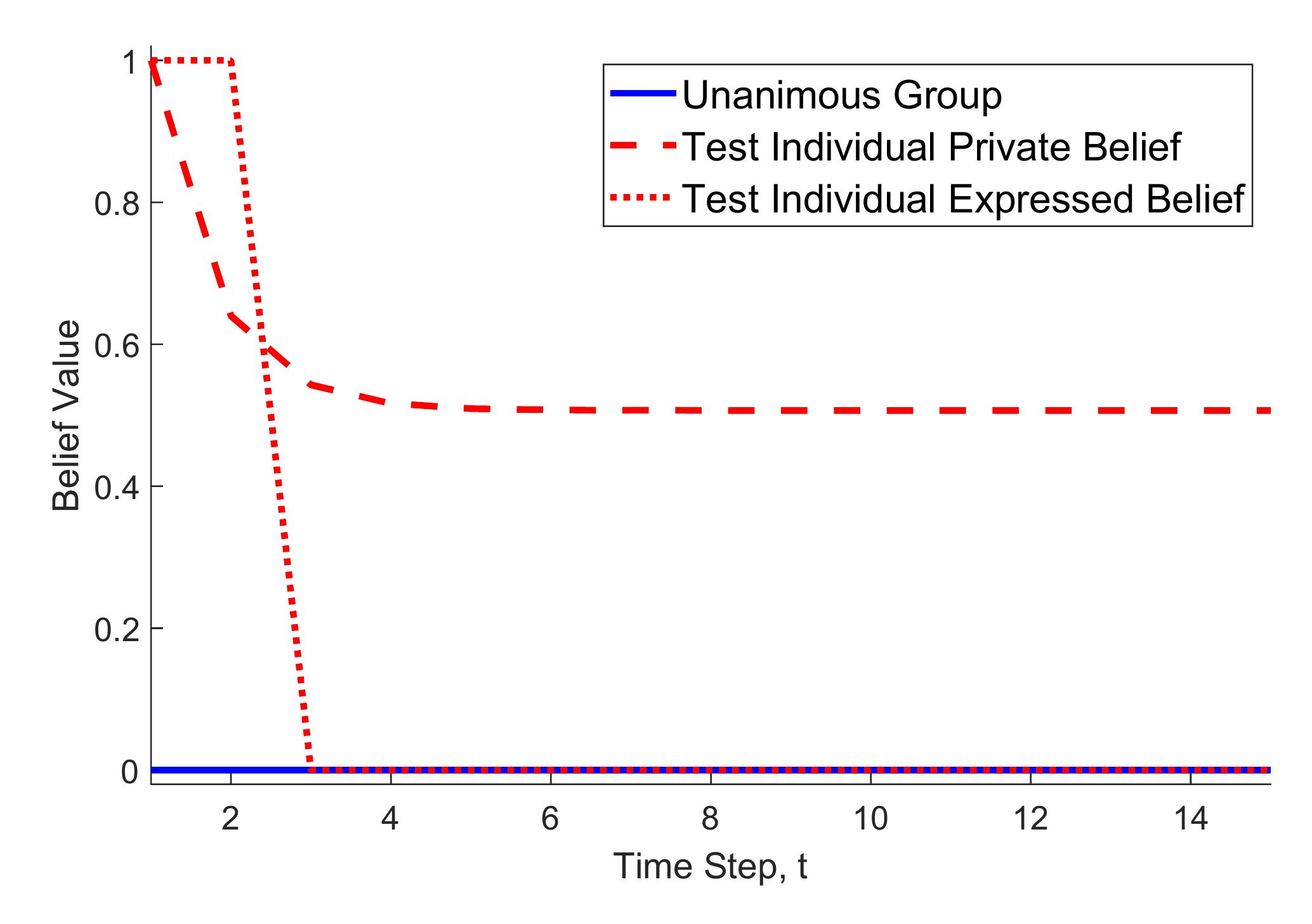}
	\caption{Asch's second experiment: An individual with $\lambda_1 = 0.9, \phi_1 = 0.1$, threshold model, with $\tau_i = 0.6$ }
	\label{fig:distortion_judgment_v2_THdif}
\end{figure}

\textit{Case 3:} Last, we simulate a test individual that has low $\phi_1$ and high $\lambda_i$  (in the original Asch setup, these individuals were said to show distortion of judgment). Fig.~\ref{fig:distortion_judgment_v2_OG} and \ref{fig:distortion_judgment_v2_TH} show a test individual with $\lambda_1 = 0.9, \phi_1 = 0.1$, for the original and threshold model, respectively. Finally, Fig.~\ref{fig:distortion_judgment_v2_THdif} shows \textit{Case 4}, which simulates a test individual with the same parameter set of $\lambda_1 = 0.9, \phi_1 = 0.1$, but with the threshold changed from $\tau_1 = 0.5$ to $\tau_1 = 0.6$.

Whether the original model or the threshold model is used, it can been seen that introduction of an actor (confederate) telling the truth has a major impact on the belief evolution of the test individual in Case 2 and 3 (compare Fig.~\ref{fig:distortion_action} with Fig.~\ref{fig:distortion_action_v2_OG} and \ref{fig:distortion_action_v2_TH}, and Fig.~\ref{fig:distortion_judgement} with Fig.~\ref{fig:distortion_judgment_v2_OG}, \ref{fig:distortion_judgment_v2_TH} and \ref{fig:distortion_judgment_v2_THdif}). The impact is significantly more pronounced under the threshold model, such that a test individual with $\lambda_1 = 0.9, \phi_1 = 0.1$ and $\tau_i = 0.5$ (Case 3) will still pick the correct answer when another actor tells the truth. When the threshold is adjusted to $\tau_i = 0.6$ (Case 4), the test individual picks the wrong answer along with the confederates. 

\section{Conclusions}\label{sec:con}
We have proposed a novel agent-based model of opinion evolution on interpersonal influence networks, where each individual has separate expressed and private opinions that evolve in a coupled manner. 
%The model describing the dynamics of the private opinion is inspired by the empirically validated Friedkin--Johnsen model, while the expressed opinion is assumed to be altered from the private opinion due to a pressure to conform to the public opinion. 
Conditions on the network and the values of susceptibility and resilience for the individuals were established for ensuring that the opinions converged exponentially fast to a steady-state of persistent disagreement. Further analysis of the final opinion values yielded semi-quantitative conclusions that led to insightful social interpretations, including the conditions that lead to a discrepancy between the expressed and private opinions of an individual. We then used the model to study Asch's experiments \cite{asch1951group_pressure_effects}, showing that all 3 types of reactions from the test individual could be captured within our framework. 
%This was achieved by attributing to the test individual intuitively reasonable values of susceptibility and resilience, underlining the generality and strength of the proposed model as a theoretical framework for investigating discrepancies in private and expressed opinions in interpersonal influence networks. Although existing static models on conformity are able to capture Asch's experiments, they cannot be generalised to arbitrary social networks. On the other hand, most opinion dynamics models cannot fully capture Asch's experiments, or other phenomena involving expressed and private opinions because a single opinion per individual per topic is assumed. Our  model thus unifies both sets of works, viz. static models of conformity and opinion dynamics models, under a single framework.
A number of interesting future directions can be considered. Preliminary simulations show that our model can also capture \emph{pluralistic ignorance}, with network structure and placement of extremist nodes having a significant effect on the observed phenomena. Clearly the threshold model in Section~\ref{ssec:threshold} requires further study, and	one could also consider the model in a continuous-time setting, or with asynchronous updating, or both.
%	, e.g. resilience may decrease as an individual's private opinion moves further from the public opinion. 

%Another promising approach is to use an event-based control framework \cite{heemels2012introduction_event} to study the ``spiral of silence'' \cite{noelle1993spiral,taylor1982pluralistic_sos}. 

%Last, we believe that the proposed model can be used to guide and give insight into designing and implementing future experiments or field studies investigating how discrepancies in private and expressed opinions arise and evolve in more general social situations than that encountered in Asch's experiments, such as online social media networks. If the empirical data obtained is in line with the model predictions, then this provides a strong validation for the model. On the other hand, a mismatch between empirical data and the model predictions may suggest unmodelled aspects of relevant human behaviour that should be incorporated to further strengthen the proposed model. This approach may help build a truly robust framework for the study of how discrepancies in private and expressed opinions arise.

\begin{ack}                               % Place acknowledgements
The authors would like to thank Julien Hendricks for his helpful discussion on the proof of Theorem~\ref{thm:PE_disagree}, and the reviewers and editor who improved the manuscript immeasurably with their suggestions and comments.
\end{ack}

\bibliographystyle{IEEEtran}        % Include this if you use bibtex 
\bibliography{MYE_ANU}           % and a bib file to produce the 
                                 % bibliography (preferred). The
                                 % correct style is generated by
                                 % Elsevier at the time of printing.

 \appendix
\section{Preliminaries}

In this section, we record some definitions, and notations to be used in the proofs of the main results. A square matrix $\mat{A} \geq 0$ is primitive if there exists $k \in \mathbb{N}$ such that $\mat{A}^k > 0$ \cite[Definition 1.12]{bullo2009distributed}. A graph $\mathcal{G}[\mat{A}]$ is strongly connected and aperiodic if and only if $\mat{A}$ is primitive, i.e. $\exists k\in \mathbb{N}$ such that $\mat{A}^k$ is a positive matrix \cite[Proposition 1.35]{bullo2009distributed}. We denote the $i^{th}$ canonical base unit vector of $\mathbb{R}^n$ as $\mathbf{e}_i$. The spectral radius of a matrix $\mat{A}\in\mathbb{R}^{n\times n}$ is given by $\rho(\mat{A})$. 

%\begin{definition}[{\cite[Definition 1.12]{bullo2009distributed}}]
%	A square matrix $\mat{A} \geq 0$ is primitive if there exists $k \in \mathbb{N}$ such that $\mat{A}^k > 0$.
%\end{definition}
%
%\begin{lemma}[{\cite[Proposition 1.35]{bullo2009distributed}}]\label{lem:primitive}
%	The graph $\mathcal{G}[\mat{A}]$ is strongly connected and aperiodic if and only if $\mat{A}$ is primitive, i.e. $\exists k\in \mathbb{N}$ such that $\mat{A}^k$ is a positive matrix.
%\end{lemma}

\begin{lemma}\label{lem:spectrum_irre}
	If $\mat{A}\in \mathbb{R}^{n\times n}$ is row-substochastic and irreducible, then $\rho (\mat{A})<1$.
\end{lemma}
\emph{\textbf{Proof}:}
This lemma is an immediate consequence of \cite[Lemma 2.8]{varga2009matrix_book}.\hfill $\qed$

%\begin{lemma}[{\cite[pg. 108--109]{bernstein2009matrixbook}}]\label{lem:block_inverse}
%	Suppose that an invertible matrix $\mat{Z} \in \mathbb{R}^{n\times n}$ is partitioned as
%	\begin{equation}
%	\mat{Z} = \begin{bmatrix} \mat{A} & \mat{B} \\ \mat{C} & \mat{D} \end{bmatrix}
%	\end{equation}
%	Suppose further that $\mat A$, $\mat{D}$ are invertible. Then $\mat{E} = \mat D - \mat{CA}^{-1}\mat{B}$ and $\mat{F} = \mat{A} - \mat{BD}^{-1}\mat{C}$ are invertible, and the inverse of $\mat{Z}$ is given as
%	\begin{equation}
%	\mat{Z}^{-1} = \begin{bmatrix} \mat{F}^{-1} & -\mat{A}^{-1}\mat B\mat{E}^{-1} \\ - \mat{D}^{-1}\mat{C}\mat{F}^{-1} & \mat{E}^{-1} \end{bmatrix}
%	\end{equation}
%\end{lemma}

%\begin{lemma}[{\cite[Fact 10.11.20]{bernstein2009matrixbook}}]\label{lem:matrix_diff}
%	Consider a matrix function $\mat{A}(s) \in \mathbb{R}^{n\times n}$, where $s \in \mathcal{D} \subseteq \mathbb{R}$. Suppose that $\mat{A}(s)$ is continuously differentiable, and invertible, for all $s\in \mathcal{D}$. Then 
%	\begin{equation}
%	\frac{\mathrm{d} \mat{A}^{-1}(s)}{\mathrm{d} s} = -\mat{A}^{-1}(s) \left(\frac{\mathrm{d} \mat{A}(s)}{\mathrm{d} s}\right)\mat{A}^{-1}(s).
%	\end{equation}
%\end{lemma}

\subsection{Performance Function and Ergodicity Coefficient}\label{ssec:app_scramble}
In order to analyse the disagreement among the opinions at steady state, we introduce a performance function and a coefficient of ergodicity. For a vector $\vect{x}\in\mathbb{R}^n$, define the performance function $V(\vect{x}):\mathbb{R}^n \mapsto \mathbb{R}$ as 
\begin{equation}\label{eq:V_function}
V(\vect{x})=\max_{i\in \{1, \hdots, n\}}x_i-\min_{j \in \{1, \hdots, n\}} x_j, 
\end{equation}
In context,  $V(\vect{y})$ measures the ``level of disagreement'' in the vector of opinions $\vect{y}(t)$, and consensus of opinions, i.e. $\vect{y}(t) = \alpha \vect{1}_n, \alpha \in \mathbb{R}$, is reached if and only if $V(\vect{y}(t))=0$. Next consider the following coefficient of ergodicity, $\tau(\mat{A})$ for a row-stochastic matrix $\mat{A}\in \mathbb{R}^{n\times n}$, defined \cite{seneta2006non} as
\begin{align}\label{eq:coeffi_ergo}
\tau \left( \mat{A} \right) & = 1-\min \limits_{i,j \in \{1, \hdots, n\}}\sum\limits_{s=1}^n \min \{a_{is},a_{js}\}.
\end{align}
This coefficient of ergodicity satisfies $0\le \tau(\mat{A}) \le1$, and $\tau(\mat{A})= 0$ if and only if $\mat{A} = \vect{1}_n\vect{z}^\top$ for some $\vect{z}\geq 0$. Importantly, there holds $\tau(\mat{A})< 1$ if $\mat{A} > 0$. Also, there holds $V(\mat{Ax})\le\tau(\mat{A})V(\mat{x})$ (see \cite{seneta2006non})

\subsection{Supporting Lemmas}

 Two lemmas are introduced to establish several properties of $\mat{P}$ and $(\mat{I}_{2n} - \mat{P})^{-1}$, which will be used to help prove the main results.
 
 \begin{lemma}\label{lem:PE_P_result}
	Suppose that Assumption~\ref{assm:PE_network} holds. Then, $\mat{P}$ given in \eqref{eq:P_matrix} is nonnegative, the graph $\mathcal{G}[\mat{P}]$ is strongly connected and aperiodic, and there holds $\rho(\mat{P}) < 1$.
\end{lemma}
 
 \begin{lemma}\label{lem:PE_Q_properties}
 	Suppose that Assumption~\ref{assm:PE_network} holds. With $\mat{P}$ given in \eqref{eq:P_matrix}, define $\mat{Q}$ as
 	\begin{equation*}
 	\mat{Q} = \begin{bmatrix} \mat{Q}_{11} & \mat{Q}_{12} \\ \mat{Q}_{21} & \mat{Q}_{22} \end{bmatrix} = \begin{bmatrix} \mat{I}_n - \mat{P}_{11} & - \mat{P}_{12} \\ -\mat{P}_{21} & \mat{I}_n - \mat{P}_{22} \end{bmatrix}.
 	\end{equation*}
 	Then, $\mat{Q}_{11}, \mat{Q}_{22}$ are nonsingular, and $\mat{Q}^{-1} >0$ is
 	\begin{equation}\label{eq:Q_inverse}
 	\mat{Q}^{-1} = \begin{bmatrix} \mat{A}& \mat{B} \\ \mat{C} & \mat{D} \end{bmatrix},
 	\end{equation}
 	where $\mat{A} =  (\mat{Q}_{11} - \mat{Q}_{12}\mat{Q}_{22}^{-1} \mat{Q}_{21})^{-1}$, $\mat{D} = (\mat{Q}_{22} - \mat{Q}_{21} \mat{Q}_{11}^{-1} \mat{Q}_{12} )^{-1}$, $\mat{B} = -\mat{Q}_{11}^{-1} \mat{Q}_{12} \mat{D} $, $\mat{C} = -\mat{Q}_{22}^{-1}\mat{Q}_{21}\mat{A}$. Moreover, $\mat{R} = \mat{A}(\mat{I}_n - \mat{\Lambda})$ and $\mat{S} = - \mat{Q}_{22}^{-1}\mat{Q}_{21}$ are invertible, positive row-stochastic matrices.
 \end{lemma}

\section{Proofs}

\subsection{Proof of Lemma~\ref{lem:PE_P_result}}
First, we verify that $\mat{P} \geq 0$ by using the fact that $\mat{W}$, $\mat{\Lambda}$, $\mat{I}_n - \mat{\Phi}$, $\mat M$ are all nonnegative.
%(since $0< \phi_i, \lambda_i <1$). 
Next, observe that
\begin{align*}
& \begin{bmatrix}
\mat{\Lambda}( \wt{\mat{W}} + \wh{\mat{W}}\mat{\Phi} ) \hspace{6pt} & \mat{\Lambda}\wh{\mat{W}}\left(\mat{I}_n-\mat{\Phi}  \right)\mat M \\
\mat{\Phi} & \left( \mat{I}_n - \mat{\Phi}\right) \mat M
\end{bmatrix} \begin{bmatrix} \vect{1}_n \\ \vect{1}_n \end{bmatrix}   = \begin{bmatrix} \mat{\Lambda}\vect{1}_n \\ \vect{1}_n \end{bmatrix} 
%\label{eq:row_sum_proof}
\end{align*}
%\begin{align}
%& \begin{bmatrix}
%\mat{\Lambda}( \wt{\mat{W}} + \wh{\mat{W}}\mat{\Phi} ) & \mat{\Lambda}\wh{\mat{W}}\left(\mat{I}_n-\mat{\Phi}  \right)\frac{\vect{1}_n \vect{1}_n^\top}{n} \\
%\mat{\Phi} & \left( \mat{I}_n - \mat{\Phi}\right) \frac{\vect{1}_n\vect{1}_n^\top }{n}
%\end{bmatrix} \begin{bmatrix} \vect{1}_n \\ \vect{1}_n \end{bmatrix}  \nonumber \\
%& \quad = 
%\begin{bmatrix} \mat{\Lambda}( \wt{\mat{W}} + \wh{\mat{W}}\mat{\Phi} )\vect{1}_n + \mat{\Lambda}\wh{\mat{W}}\left(\mat{I}_n-\mat{\Phi}  \right)\vect{1}_n \\ \mat{\Phi}\vect{1}_n + \left( \mat{I}_n - \mat{\Phi}\right) \vect{1}_n  \end{bmatrix} \nonumber \\
%& \quad = \begin{bmatrix} \mat{\Lambda}(\wt{\mat{W}} + \wh{\mat{W}})\vect{1}_n \\ \vect{1}_n \end{bmatrix}  = \begin{bmatrix} \mat{\Lambda}\vect{1}_n \\ \vect{1}_n \end{bmatrix}. \label{eq:row_sum_proof}
%\end{align}
because $\mat M$ and $\mat{W} = \wt{\mat{W}}+\wh{\mat{W}}$ are row-stochastic. 

Notice that the graph $\mathcal{G}[\mat{P}] = (\mathcal{V},\mathcal{E}[\mat{P}], \mat{P})$ has $2n$ nodes, with $\mathcal{V} =\{1, \hdots, 2n\}$. The node subset $\mathcal{V}_1 = \{v_1,\hdots,v_n\}$ contains node $v_i$ which is associated with individual $i$'s private opinion $y_i$, $i \in \mathcal{I}$. The node subset $\mathcal{V}_2 = \{v_{n+1}, \hdots, v_{2n}\}$ contains node $v_{n+i}$ which is associated with individual $i$'s expressed opinion $\hat{y}_i$, $i\in\mathcal{I}$. Define the following two subgraphs; $\mathcal{G}_1 = (\mathcal{V}_1, \mathcal{E}[\mat{P}_{11}], \mat{P}_{11})$ and $\mathcal{G}_2 = (\mathcal{V}_2, \mathcal{E}[\mat{P}_{22}], \mat{P}_{22})$. The edge set of $\mathcal{G}[\mat{P}]$ can be divided as follows
\begin{align*}
\mathcal{E}_{11} & = \mathcal{E}\begin{bmatrix}
\mat{P}_{11} & \mat{0}_{n\times n} \\
\mat{0}_{n\times n} & \mat{0}_{n\times n} \end{bmatrix},\quad \mathcal{E}_{12} = \mathcal{E}\begin{bmatrix}
\mat{0}_{n\times n} & \mat{P}_{12} \\
\mat{0}_{n\times n} & \mat{0}_{n\times n} \end{bmatrix} ,\quad \\
\mathcal{E}_{21} &  = \mathcal{E}\begin{bmatrix}
\mat{0}_{n\times n} & \mat{0}_{n\times n} \\
\mat{P}_{21} & \mat{0}_{n\times n} \end{bmatrix}  ,\quad
\mathcal{E}_{22} = \mathcal{E}\begin{bmatrix}
\mat{0}_{n\times n} & \mat{0}_{n\times n} \\
\mat{0}_{n\times n} & \mat{P}_{22} \end{bmatrix} ,
\end{align*}
% \begin{align}
% \mathcal{E}_{11} & = \mathcal{E}\begin{bmatrix}
% \mat{P}_{11} & \mat{0}_{n\times n} \\
% \mat{0}_{n\times n} & \mat{0}_{n\times n} \end{bmatrix} \\
% \mathcal{E}_{12} & = \mathcal{E}\begin{bmatrix}
% \mat{0}_{n\times n} & \mat{P}_{12} \\
% \mat{0}_{n\times n} & \mat{0}_{n\times n} \end{bmatrix} \\
% \mathcal{E}_{21} & = \mathcal{E}\begin{bmatrix}
% \mat{0}_{n\times n} & \mat{0}_{n\times n} \\
% \mat{P}_{21} & \mat{0}_{n\times n} \end{bmatrix}  \\
% \mathcal{E}_{22} & = \mathcal{E}\begin{bmatrix}
% \mat{0}_{n\times n} & \mat{0}_{n\times n} \\
% \mat{0}_{n\times n} & \mat{P}_{22} \end{bmatrix} 
% \end{align}
In other words, $\mathcal{E}_{11}$ contains only edges between nodes in $\mathcal{V}_1$ and $\mathcal{E}_{22}$ contains only edges between nodes in $\mathcal{V}_2$. The edge set $\mathcal{E}_{12}$ contains only edges from nodes in $\mathcal{V}_2$ to nodes in $\mathcal{V}_1$, while the edge set $\mathcal{E}_{21}$ contains only edges from nodes in $\mathcal{V}_1$ to nodes in $\mathcal{V}_2$. Clearly $\mathcal{E}[\mat{P}] = \mathcal{E}_{11}\cup \mathcal{E}_{12} \cup \mathcal{E}_{21}\cup \mathcal{E}_{22}$. It will now be shown that $\mathcal{G}[\mat{P}]$ is strongly connected and aperiodic, implies that $\mat{P}$ is primitive.

Since the diagonal entries of $\mat\Lambda, \mat\Phi$ are strictly positive, it is obvious that $\mat{P}_{11} = \mat{\Lambda}( \wt{\mat{W}} + \wh{\mat{W}}\mat{\Phi} ) \sim \mat{W}$. Because $\mathcal{G}[\mat{W}]$ is strongly connected and aperiodic, it follows that $\mathcal{G}_1$ is strongly connected and aperiodic. Similarly, the edges of $\mathcal G_2$ are $\mathcal{E}[\mat{P}_{22}]$. Because $\mat{I}_n-\mat\Phi$ has strictly positive diagonal entries, one concludes that $\mat{P}_{22} = \left( \mat{I}_n - \mat{\Phi}\right) \mat M \sim \mathcal{G}[\mat M] \sim \mathcal{G}[\mat W]$, i.e. $\mathcal{G}_2$ is strongly connected and aperiodic. Since $\mathcal{G}_1$ and $\mathcal{G}_2$ are both, separately, strongly connected, then if there exists 1) an edge from any node in $\mathcal V_1$ to any node $\mathcal V_2$, and 2) an edge from any node in $\mathcal V_2$ to any node in $\mathcal V_1$, one can conclude that the graph $\mathcal G[\mat{P}]$ is strongly connected. It suffices to show that $\mathcal E_{12} \ne \emptyset$ and $\mathcal E_{21} \ne \emptyset$. Since $\mat{P}_{21} = \mat{\Phi}$ has strictly positive diagonal entries, this proves that $\mathcal E_{12} \ne \emptyset$. From the fact that $\mat{I}_n-\mat{\Phi}$ has strictly positive diagonal entries, and because $\wh{\mat{W}}$ is irreducible, it follows that $\mat{P}_{12} = \mat{\Lambda}\wh{\mat{W}}\left(\mat{I}_n - \mat{\Phi} \right)\mat M \ne \mat 0_{n\times n}$. This shows that $\mathcal E_{21} \ne \emptyset$. 

It has therefore been proved that $\mathcal{G}[\mat{P}]$ is strongly connected and aperiodic, which also proves that $\mat{P}$ is irreducible. 
%Because $\lambda_i < 1\,\forall\,i$, it is immediately clear that \eqref{eq:row_sum_proof} implies that rows $1, \hdots, n$ of $\mat{P}$ each have row sum equal to a value strictly less than one, while rows $n+1, \hdots, 2n$ each have row sum precisely equal to one. 
Since $\lambda_i < 1\,\forall\,i$, $\mat{P}$ is row-substochastic, Lemma~\ref{lem:spectrum_irre} establishes that $\rho(\mat{P}) < 1$.
%, i.e. all eigenvalues of $\mat{P}$ are inside the unit circle. 
This completes the proof. 
\hfill $\square$

\subsection{Proof of Lemma~\ref{lem:PE_Q_properties}}
Lemma~\ref{lem:PE_P_result} showed that $\mathcal{G}[\mat{P}]$ is strongly connected and aperiodic, which implies that $\mat{P}$ is primitive. Since $\mat{Q}^{-1} = (\mat{I}_{2n} - \mat{P})^{-1}$ ans $\rho(\mat{P}) < 1$, the Neumann series yields $\mat{Q}^{-1} =  \sum_{k = 0}^{\infty} \mat{P}^k > 0$. Next, it will be shown $\mat{Q}_{11}$, $\mat{Q}_{22}$ and $\mat{D} = \mat{Q}_{11} - \mat{Q}_{12} \mat{Q}_{22}^{-1} \mat{Q}_{21}$ are all invertible, which will allow $\mat{Q}^{-1}$ to be expressed in the form of \eqref{eq:Q_inverse}  by use of \cite[Proposition 2.8.7, pg. 108--109]{bernstein2009matrixbook}. Under Assumption~\ref{assm:PE_network}, $\mathcal{G}_1[\mat{P}_{11}]$ and $\mathcal{G}_2[\mat{P}_{22}]$ are both strongly connected and aperiodic; Lemma~\ref{lem:spectrum_irre} states that $\rho(\mat{P}_{11}), \rho(\mat{P}_{22}) < 1$. Since $\mat{Q}_{11} =  \mat{I}_n - \mat{P}_{11}$ and $\mat{Q}_{22} = \mat{I}_n - \mat{P}_{22}$, the same method as above can be used to prove that $\mat{Q}_{11}, \mat{Q}_{22}$ are invertible, and satisfy $\mat{Q}^{-1}_{11}, \mat{Q}^{-1}_{22} > 0$.

In order to prove that $\mat{D}$ is invertible, we first establish some properties of $\mat{S} = - \mat{Q}_{22}^{-1} \mat{Q}_{21}$. Since $\mat{Q}_{22}^{-1} > 0$, it follows from the fact that $\mat{\Phi} = \text{diag}(\phi_i)$ is a positive diagonal matrix, that $\mat{S} = \mat{Q}_{22}^{-1} \mat{\Phi} > 0$. To prove that $\mat{S}$ is row-stochastic, first note that $\det(\mat{S}) =\det(\mat{Q}_{22}^{-1})\det(\mat{\Phi}) \neq 0$  (we have $\phi_i \in (0,1), \forall\, i \Rightarrow \det(\mat{\Phi}) \neq 0$). Since $(\mat{AB})^{-1} = \mat{B}^{-1} \mat{A}^{-1}$, observe that
\begin{align}
\mat{S} = \big( \mat{\Phi}^{-1} - \mat{\Phi}^{-1} (\mat{I}_n - \mat{\Phi}) \mat M \big)^{-1}. \label{eq:S_inv}
\end{align}
From \eqref{eq:S_inv}, verify that $\mat{S}^{-1}\vect{1}_n = \vect{1}_n$, which implies
%\begin{align*}
%\mat{S}^{-1}\vect{1}_n & = \big( \mat{\Phi}^{-1} - \mat{\Phi}^{-1} (\mat{I}_n - \mat{\Phi}) \frac{\vect{1}_n \vect{1}_n^\top }{n} \big)\vect{1}_n  = \vect{1}_n.
%\end{align*}
 $\mat{S}\mat{S}^{-1}\vect{1}_n = \mat{S}\vect{1}_n \Leftrightarrow \mat{S}\vect{1}_n = \vect{1}_n$, i.e. $\mat{S}$ is row-stochastic.

We now turn to proving that $\mat{D}$ is invertible. Notice that $\mat{S}$, $-\mat{Q}_{12} = \mat{P}_{12}$, and $\mat{\Lambda}(\wt{\mat W} + \wh{\mat W}\mat{\Phi})$ are all nonnegative. We write $\mat{D} = \mat{I}_n - \mat{U}$ where $\mat{U} = \mat P_{11} + \mat{P}_{12}\mat{S}\geq 0 $. Observe that $\mat{U}\vect{1}_n  =\mat P_{11}\vect{1}_n + \big(\mat{\Lambda}\wh{\mat{W}}\left(\mat{I}_n-\mat{\Phi}  \right)\big)\vect{1}_n  = \mat{\Lambda}\vect{1}_n$
%\begin{align*}
%& \mat{U}\vect{1}_n  =\mat P_{11}\vect{1}_n + \big(\mat{\Lambda}\wh{\mat{W}}\left(\mat{I}_n-\mat{\Phi}  \right)\big)\vect{1}_n  = \mat{\Lambda}\vect{1}_n
%\end{align*}
because $(\wh{\mat W} + \wt{\mat W})\vect{1}_n = \vect{1}_n$. In other words, the $i^{th}$ row of $\mat{U}$ sums to $\lambda_i < 1$ (see Assumption~\ref{assm:PE_network}), which implies that $\Vert \mat{U} \Vert_{\infty} < 1 \Rightarrow \rho(\mat{U}) < 1$. Because it was shown in the proof of Lemma~\ref{lem:PE_P_result} that $\mathcal{G}[\mat P_{11}]$ is strongly connected and aperiodic, it is straightforward to show that $\mathcal{G}[\mat{U}]$ is also strongly connected and aperiodic. It follows that $\mat{U}$ is primitive, which implies that $\mat{D}^{-1} > 0$ from the Neumann series $\mat{D}^{-1} = \sum_{k=0}^\infty \mat{U}^k$.  Thus, $\mat{R} = \mat{D}^{-1}(\mat{I}_n - \mat{\Lambda}) > 0$, because $\mat{I}_n - \mat{\Lambda}$ is a positive diagonal matrix. Finally, one can verify that $\mat{R}$ is row-stochastic with the following computation: $\mat{D}\vect{1}_n = (\mat{I}_n - \mat{U})\vect{1}_n =  (\mat{I}_n - \mat{\Lambda})\vect{1}_n \Rightarrow \mat{R}\vect{1}_n = \mat{D}^{-1}(\mat{I}_n - \mat{\Lambda})\vect{1}_n = \mat{D}^{-1} \mat{D}\vect{1}_n = \vect{1}_n$. This completes the proof. \hfill $\square$

\subsection{Proof of Theorem~\ref{thm:PE_convergence_eqb} and Corollary~\ref{cor:PE_consensus}}
%	The proofs make extensive use of Lemma~\ref{lem:PE_P_result} and \ref{lem:PE_Q_properties}.

\emph{Proof of Theorem \ref{thm:PE_convergence_eqb}:} Lemma~\ref{lem:PE_P_result} established that the time-invariant matrix $\mat{P}$ satisfies $\rho(\mat{P}) < 1$. Standard linear systems theory \cite{rugh1996linearsystems_book} is used to conclude that the linear, time-invariant system \eqref{eq:system_compact}, with constant input $\left[((\mat{I}_n - \mat{\Lambda})\vect{y}(0))^\top, \; \vect{0}_n^\top \right]^\top$, converges exponentially fast to 
\begin{align}
\begin{bmatrix} \lim_{t\to\infty} \vect{y}(t) \\ \lim_{t\to\infty} \hat{\vect y}(t) \end{bmatrix}  \triangleq \begin{bmatrix} \vect{y}^*\\ \hat{\vect y}^* \end{bmatrix} & = (\mat{I}_{2n} -\mat{P})^{-1} \! \begin{bmatrix} (\mat{I}_n - \mat{\Lambda})\vect{y}(0) \\ \vect{0}_n \end{bmatrix} \nonumber \\
&  = \mat{Q}^{-1} \begin{bmatrix} (\mat{I}_n - \mat{\Lambda})\vect{y}(0) \\ \vect{0}_n \end{bmatrix}.
\end{align}
Having calculated the form of $\mat{Q}^{-1}$ in \eqref{eq:Q_inverse}, it is straightforward to verify that $\vect{y}^* = \mat{R} \vect{y}(0)$ and $\hat{\vect y}^* = \mat{S} \mat{R} \vect{y}(0) = \mat{S} \vect{y}^*$. Here, the definitions of $\mat{R}$ and $\mat{S}$ are given in Lemma~\ref{lem:PE_Q_properties}, which also proved their positivity and row-stochasticity. This completes the proof. \hfill $\qed$

\emph{Proof of Corollary~\ref{cor:PE_consensus}:} The assumption that $\mat{\Lambda} = \mat{I}_n$ implies that $\mat{P}$ is nonnegative and row-stochastic. The proof of Lemma~\ref{lem:PE_P_result} established that $\mathcal{G}[\mat{P}]$ is strongly connected and aperiodic, and this remains unchanged when $\mat{\Lambda} = \mat{I}_n$. Standard results on the DeGroot model \cite{proskurnikov2017tutorial} then imply that consensus is achieved exponentially fast, i.e. $\lim_{t\to\infty} \vect{y}(t) = \hat{\vect y}(t) = \alpha \vect{1}_{n}$ for some $\alpha \in \mathbb{R}$. \hfill $\qed$

\subsection{Proof of Theorem~\ref{thm:PE_disagree}}
If $\vect{y}(0) = \alpha \vect{1}_n$, for some $\alpha\in \mathbb{R}$ (i.e. the initial private opinions are at a consensus), then $\vect{y}^* = \hat{\vect y}^* = \alpha \vect{1}_n$ because $\mat{R}$ and $\mat{S}$ are row-stochastic. In what follows, it will be proved that if the initial private opinions are not at a consensus, then there is disagreement at steady state. 
%It is suggested that the reader become familiar with the performance function $V(\vect{x})$ and coefficient of ergodicity $\tau(\mat{A})$ in Appendix~\ref{ssec:app_scramble}, as these will be used frequently in this proof.

First, we establish $y_{\min}^* \neq y_{\max}^*$. Note that $V(\vect{y}^*) = 0$ if and only if $\vect{y}^* = \beta \vect{1}_n$, for some $\beta \in \mathbb{R}$. Next, observe that $\vect{y}^* = \beta \vect{1}_n$ if and only if $\mat{R}\vect{y}(0) = \beta \vect{1}_n$, for some $\beta \in \mathbb{R}$. Note that $\mat{R}$ is invertible, because it is the product of two invertible matrices (see Lemma~\ref{lem:PE_Q_properties}). Moreover, because $\mat{R}$ is row-stochastic, there holds $\mat{R}\vect{1}_n = \vect{1}_n \Leftrightarrow \mat{R}^{-1} \mat{R}\vect{1}_n = \mat{R}^{-1}\vect{1}_n \Leftrightarrow \mat{R}^{-1}\vect{1}_n = \vect{1}_n$. Thus, premultiplying by $\mat{R}^{-1}$ on both sides of $\mat{R}\vect{y}(0) = \beta \vect{1}_n$ yields $\vect{y}(0) = \beta \mat{R}^{-1}\vect{1}_n = \beta \vect{1}_n$. In other words, a consensus of the final private opinions, $\vect{y}^* = \beta \vect{1}_n $, occurs if and only if the initial private opinions are at a consensus. Recalling the theorem hypothesis that $\vect{y}(0) \neq \alpha \vect{1}_n$, for some $\alpha \in \mathbb{R}$, it follows that $\vect{y}^*$ is not at a consensus. Thus, $y_{\min}^* \neq y_{\max}^*$ as claimed.

Next, the inequalities \eqref{eq:max_ineq} and \eqref{eq:min_ineq} are proved. Since $\mat{R}, \mat{S} > 0$ are row-stochastic, $\tau(\mat{R}), \tau(\mat{S}) < 1$. Because $\mat{R}$ is invertible, $\mat{R} \neq \vect{1}_n \vect{z}^\top$ for some $\vect{z} \in \mathbb{R}^n$. This means that $\tau(\mat{R}) > 0$ (see below \eqref{eq:coeffi_ergo}). Similarly, one can prove that $\tau(\mat{S}) > 0$. In the above paragraph, it was shown that if there is no consensus of the initial private opinions, then $V(\vect{y}^* = \mat{R}\vect{y}(0)) > 0$. By recalling that $V(\mat{Ax}) \leq \tau(\mat{A})V(\vect{x})$ (see Appendix~\ref{ssec:app_scramble}) and the above facts, we conclude that $0 < V(\vect{y}^* = \mat{R}\vect{y}(0)) < V(\vect{y}(0))$, which establishes the left hand inequality of \eqref{eq:max_ineq} and \eqref{eq:min_ineq}. Following steps similar to the above, but which are omitted, one can show that  $0 < V(\hat{\vect{y}}^* = \mat{S}\vect{y}^*) < V(\vect{y}^*)$, which establishes the right hand inequality of \eqref{eq:max_ineq} and \eqref{eq:min_ineq}, and also establishes that $\hat{y}_{\min}^* \neq \hat{y}_{\max}^*$.

Last, it remains to prove that for generic initial conditions, $y_i^* \neq \hat{y}_i^*$. Observe that $\hat{y}_i^* = y_i^* \Rightarrow \hat{y}_{avg}^* = \vect{1}_n^\top \hat{\vect y}^*/n$. Thus, $\hat{y}_i^* = y_i^*$ for $m$ specific individuals if and only if there are $m$ independent equations satisfying $(\mathbf{e}_i - \frac{1}{n}\vect 1_n)^\top \vect y^* = 0$. This implies that $\hat{y}^*$ must lie in an $n-m$-dimensional subspace of $\mathbb{R}^n$, denoted as $\mathcal{D}$. From Theorem~\ref{thm:PE_convergence_eqb}, one has $\vect{y}^* = \mat{RSy}(0)$. It follows that $\hat{y}_i^* = y_i^*$ for $m$ specific individuals only if $\vect y(0)$ belongs to the inverse image (by $\mat{RS}$) of $\mathcal{D}$, and the inverse image has dimension $n-m$ because $\mat{R}, \mat{S}$ are invertible. This completes the proof. \hfill $\qed$

\subsection{Proof of Corollary~\ref{cor:scrambling_constant}}\label{ssec:PEModel_app_est}
Recall the definition of $V$ in Appendix~\ref{ssec:app_scramble}. From Theorem~\ref{thm:PE_convergence_eqb}, one has that $V(\hat{\vect{y}}^*) = V(\mat{S}\vect{y}^*) \leq \tau(\mat{S}) V(\vect{y}^*)$, which implies that there holds \mbox{$V(\hat{\vect{y}}^*)/\tau(\mat{S}) \leq V(\vect{y}^*)$}. Thus, \eqref{eq:scrambling_bound} can be proved by showing that $\tau(\mat{S}) \leq \kappa(\vect{\phi})$. Note that since global public opinion $\hat y_{\text{avg}}$ is used, $\mat M$ in \eqref{eq:P_matrix} becomes $\mat M = n^{-1}\vect{1}_n \vect{1}_n^\top$. Recall that $\mat{Q}_{22}^{-1}$ can be expressed as $\mat{Q}_{22}^{-1} = \sum_{k=0}^\infty \mat{P}_{22}$. Since $\mat{P}_{22} = n^{-1}\left( \mat{I}_n -\mat{\Phi} \right)\vect{1}_n \vect{1}_n^\top$ and $\mat{Q}_{21} = - \mat{\Phi}$, we obtain $\mat{S}  = \mat{\Phi} + \mat H$ where $\mat H \triangleq \sum_{k=1}^\infty \big[ \left( \mat{I}_n -\mat{\Phi} \right)\frac{\vect{1}_n \vect{1}_n^\top}{n} \big]^k \mat{\Phi} > 0$.
%\begin{align*}
%\mat{S} &  = \mat{\Phi} + \sum_{k=1}^\infty \Big[\left( \mat{I}_n -\mat{\Phi} \right)\frac{\vect{1}_n \vect{1}_n^\top}{n}\Big]^k \mat{\Phi}.
%\end{align*}

Let $\underline{a} = \min_{i,j} a_{ij}$ denote the smallest element of a matrix $\mat{A}$, and 
%\begin{equation}\label{eq:scrmb_const_bound}
%\tau(\mat{A}) \leq 1 - n \underline{a}
%\end{equation}
%Define $\mat{H} = \sum_{k=1}^\infty [( \mat{I}_n -\mat{\Phi} )\frac{\vect{1}_n \vect{1}_n^\top}{n}]^k \mat{\Phi}$ and verify easily that $\mat{H} > 0$. 
observe that $\underline{s} = \underline{h}$ because $\mat{S} = \mat{\Phi} + \mat{H}$ has the same offdiagonal entries as $\mat{H}$, and the $i^{th}$ diagonal entry of $\mat{S}$ is greater than that of $\mat{H}$ by $\phi_i > 0$. Since $\mat{S}>0$, \eqref{eq:coeffi_ergo} yields $\tau(\mat{S}) \leq 1 - n \underline{s} \leq 1 - n\underline{h}$. We now analyse $\mat{H}$. For any $\mat{A} \in \mathbb{R}^{n\times n}$, there holds
\begin{align*}
& n^{-1}\left( \mat{I}_n -\mat{\Phi} \right)\vect{1}_n \vect{1}_n^\top \mat{A}  \nonumber \\
& = \frac{1}{n}
\begin{bmatrix}
(1-\phi_1)\sum_{j=1}^n a_{1j} &  \cdots & (1-\phi_1)\sum_{j=1}^n a_{nj} \\
\vdots  & \ddots & \vdots \\
(1-\phi_n)\sum_{j=1}^n a_{1j} & \cdots & (1-\phi_n)\sum_{j=1}^n a_{nj}
\end{bmatrix}.
\end{align*}
By recursion, we obtain that the $(i,j)^{th}$ entry of  $[\left( \mat{I}_n -\mat{\Phi} \right)\frac{\vect{1}_n \vect{1}_n^\top}{n}]^k$ is given by $\frac{(1-\phi_i)}{n^k} \gamma_k$, 
%\begin{equation}\label{eq:ij_expression}
%
%\end{equation}
where 
\begin{equation*}
 \gamma_k \!=\!  \Big[ {\underbrace { \sum_{p_{1}=1}^n \sum_{p_{2}=1}^n \!\!\cdots\!\!\!\! \sum_{p_{k-1}=1}^n \!\! (1-\phi_{p_1}) (1-\phi_{p_2}) \cdots (1-\phi_{p_{k-1}}) }_{\text{k-1 summation terms}} } \Big]
\end{equation*}
 This is obtained by recursively using $\sum_{i=1}^n \sum_{j=1}^n a_i b_j = \big( \sum_{i=1}^n a_i \big) \sum_{j=1}^n b_j =   \sum_{i=1}^n a_i  \big(\sum_{j=1}^n b_j\big)$.  Next, define $\mat{Z}^{k} = [\left( \mat{I}_n -\mat{\Phi} \right)\frac{\vect{1}_n \vect{1}_n^\top}{n}]^k \mat{\Phi}$. From the above, one can show that the $(i,j)^{th}$ element of $\mat{Z}^k$ is given by $z_{ij}(k) = \frac{1}{n^k}(1-\phi_i)\phi_j \gamma_k.$
%\begin{equation}\label{eq:z_ij_expression}
%
%\end{equation}
It follows that the smallest element of $\mat{Z}^k$, denoted by $\underline{z}(k)$, is bounded as follows
\begin{equation}\label{eq:z_01}
\underline{z}(k) \geq \frac{1}{n^k}(1-\phi_{\max})\phi_{\min} \gamma_k.
\end{equation}
Observe that $1-\phi_i \geq 1 - \phi_{\max},\forall\,i \Rightarrow \sum_{a=1}^n 1-\phi_a \geq n(1-\phi_{\max})$. It follows that
\begin{align}\label{eq:z_02}
\underline{z}(k) & 
%\geq \frac{1}{n^k}(1-\phi_{\max})\phi_{\min} (1-\phi_{\max})^{k-1}n^{k-1} \nonumber \\
 \geq \frac{1}{n}\phi_{\min}(1-\phi_{\max})^k.
\end{align}
Since $\mat{H} = \sum_{k=1}^\infty \mat{Z}^k$, there holds $\underline{h}  \geq \sum_{k=1}^\infty \underline{z}(k) \geq \phi_{\min}(1-\phi_{\max})(n\phi_{\max})^{-1}$.
%\begin{align*}
%\underline{h}  \geq \sum_{k=1}^\infty \underline{z}(k) 
%%& \geq n^{-1}\phi_{\min}\Big[ \sum_{k=1}^\infty (1-\phi_{\max})^k \Big]  \\
%%& \geq \frac{1}{n}\phi_{\min}\left[ \frac{1}{1-(1-\phi_{\max})} -1 \right] \\
%& \geq \phi_{\min}(1-\phi_{\max})(n\phi_{\max})^{-1},
%\end{align*}
We can obtain this by noting that for any $r \in (-1,1)$, the geometric series is $\sum_{k=0}^\infty r^k = \frac{1}{1-r} \Leftrightarrow \sum_{k=1}^\infty r^k = \frac{1}{1-r} - 1$, and $0 < 1 - \phi_{\max} < 1$. From $\tau(\mat{S}), \tau(\mat{H}) \leq 1 - n\underline{h}$, and the above arguments, we obtain $\tau(\mat{S})  \leq 1 - n \underline{h}  = 1 - \frac{\phi_{\min}}{\phi_{\max}}(1-\phi_{\max}) = \kappa(\vect{\phi})$
%\begin{align}
%\tau(\mat{S}) & \leq 1 - n \underline{h}  = 1 - \frac{\phi_{\min}}{\phi_{\max}}(1-\phi_{\max}) = \kappa(\vect{\phi})
%\end{align}
as in the corollary statement. Since $0 < \phi_{\min}/\phi_{\max} < 1$ and $0 < 1-\phi_{\max} < 1$, one has  $0 < \kappa(\vect{\phi}) < 1$ and thus $\tau(\mat{S}) \leq \kappa(\vect{\phi})$ holds $\forall\,\phi_i \in (0,1)$. \hfill $\square$

Key to the proof is that the coefficient of ergodicity for $\mat{S}$ is bounded from above as $\tau(\mat{S})\leq \kappa(\vect{\phi})$. The tightness of $\tau(\mat{S})\leq \kappa(\vect{\phi})$ depends on $\phi_{\min}/\phi_{\max}$: this can be concluded by examining the proof, and noting that the key inequalities in \eqref{eq:z_01} and \eqref{eq:z_02} involve $\phi_{\min}$ and $\phi_{\max}$. If $\phi_{\min}/\phi_{\max} = 1$, then $\tau(\mat{S}) = \kappa(\vect{\phi})$.

\subsection{Proof of Corollary~\ref{cor:derivative}}
First, verify that $\mat{S}$ is invertible, and continuously differentiable, for all $\phi_i \in (0,1)$. From \cite[Fact 10.11.20]{bernstein2009matrixbook} we obtain
\begin{equation}\label{eq:deriv_S}
\frac{\partial \mat{S}(\vect{\phi})}{\partial \phi_i} = - \mat{S}(\vect{\phi}) \left( \frac{\partial \mat{S}^{-1}(\vect{\phi})}{\partial \phi_i}\right) \mat{S}(\vect{\phi}).
\end{equation}
Below, the argument $\vect{\phi}$ will be dropped from $\mat{S}(\vect{\phi})$ and $\mat{S}^{-1}(\vect{\phi})$ when there is no confusion. Note that $\frac{\partial \mat{\Phi}^{-1}}{\partial \phi_i} = -\phi_i^{-2} \mathbf{e}_i\mathbf{e}_i^\top$. Using \eqref{eq:S_inv} and \eqref{eq:deriv_S}, one obtains $\frac{\partial \mat{S}(\vect{\phi})}{\partial \phi_i}  = \phi_i^{-2} \mat{S} \mathbf{e}_i \left( \mathbf{e}_i^\top - \vect m_i^\top \right)\mat{S}$,
%\begin{align*}
%\frac{\partial \mat{S}(\vect{\phi})}{\partial \phi_i} & = \phi_i^{-2} \mat{S} \mathbf{e}_i \left( \mathbf{e}_i^\top - \vect m_i^\top \right)\mat{S},
%\end{align*}
where $\vect m_i^\top$ is the $i^{th}$ row of $\mat M$. It suffices to prove the corollary claim, if it can be shown that the row vector $\left( \mathbf{e}_i^\top - \vect m_i^\top\right)\mat{S}$ has a strictly positive $i^{th}$ entry and all other entries are strictly negative. This is because $\mat{S} > 0 \Rightarrow \mat{S} \mathbf{e}_i > 0$. We achieve this by showing that
\begin{align}
( \mathbf{e}_i^\top - \vect m_i^\top )\mat{S}\mathbf{e}_i & > 0 \label{eq:diff_prove_01} \\
( \mathbf{e}_i^\top - \vect m_i^\top )\mat{S}\mathbf{e}_j & < 0\, ,\;\forall\, j \neq i. \label{eq:diff_prove_02}
\end{align}
Observe the following useful quantity:
\begin{align}
\mathbf{e}_i^\top \mat{S}^{-1} &  = \mathbf{e}_i^\top \big( \mat{\Phi}^{-1} - \mat{\Phi}^{-1} (\mat{I}_n - \mat{\Phi}) \mat M \big) \nonumber \\
& = \phi_i^{-1}\mathbf{e}_i^\top - ( \phi_i^{-1} - 1 ) \vect{m}_i^\top. \label{eq:phi_c}
\end{align}
%which holds because $\mathbf{e}_i^\top\mat{\Phi}^{-1} = \phi_i^{-1}\mathbf{e}_i^\top$. 
Postmultiplying by $\mat{S}$ on both sides of \eqref{eq:phi_c} yields $\mathbf{e}_i^\top = \phi_i^{-1}\mathbf{e}_i^\top \mat{S} - (\phi_i^{-1} - 1)\vect{m}_i^\top \mat{S}$.
%\begin{equation*}
%\mathbf{e}_i^\top = \frac{1}{\phi_i}\mathbf{e}_i^\top \mat{S} - \frac{1}{n}\left(\frac{1}{\phi_i} - 1\right)\vect{1}_n^\top \mat{S}.
%\end{equation*}
Rearranging this yields 
\begin{align}
\mathbf{e}_i^\top \mat{S} & = \phi_i \mathbf{e}_i^\top + (1-\phi_i)\vect{m}_i^\top \mat{S} \label{eq:eS} \\
\vect{m}_i^\top \mat{S} & = (1-\phi_i)^{-1}\left(\mathbf{e}_i^\top\mat{S} - \phi_i \mathbf{e}_i^\top \right). \label{eq:1S}
\end{align}
%First, \eqref{eq:diff_prove_02} will be proved. 
By using the equality of \eqref{eq:eS} for substitution, observe that the left hand side of \eqref{eq:diff_prove_02} is
\begin{align*}
& ( \mathbf{e}_i^\top\mat{S}  - \vect{m}_i^\top \mat{S} )\mathbf{e}_j \nonumber \\
 & \!=\! \big( \phi_i \mathbf{e}_i^\top  + (1-\phi_i)\vect{m}_i^\top \mat{S} -\vect{m}_i^\top \mat{S} \big) \mathbf{e}_j  = -\phi_i \vect{m}_i^\top \mat{S} \vect{e}_j, 
\end{align*}
because $\vect{e}_i^\top \vect{e}_j = 0$ for any $j \neq i$.  Note that $\vect{m}_i^\top \mat{S} \vect{e}_j > 0$ because $\mat M$ being irreducible implies $\vect m_i^\top \neq \vect 0_n^\top$. Thus,  $-\phi_i\vect{m}_i^\top \mat{S} \vect{e}_j/n < 0$, which proves \eqref{eq:diff_prove_02}. Substituting the equality in \eqref{eq:1S}, observe that the left hand side of \eqref{eq:diff_prove_01} is
\begin{align}
 (\mathbf{e}_i^\top \mat{S} - \vect{m}_i^\top \mat{S})& \mathbf{e}_i  = \mathbf{e}_i^\top \mat{S} \mathbf{e}_i - \frac{1}{1-\phi_i}\left(\mathbf{e}_i^\top \mat{S} \mathbf{e}_i - \phi_i \mathbf{e}_i^\top \mathbf{e}_i \right) \nonumber \\
& = \frac{\phi_i}{1-\phi_i}\left(1 - \mathbf{e}_i^\top \mat{S} \mathbf{e}_i \right) > 0. \label{eq:diff_diag}
\end{align}
The inequality is obtained by observing that 1) $\phi_i \in (0,1) \Rightarrow \phi_i/(1-\phi_i) > 0$, and 2) $1- \mathbf{e}_i^\top\mat{S} \mathbf{e}_i > 0$ because $0< \mathbf{e}_i^\top \mat{S} \mathbf{e}_i = s_{ii} < 1$. This proves \eqref{eq:diff_prove_01}. \hfill $\qed$

\balance

\end{document}